\documentclass{aa}
\usepackage{graphicx}
\usepackage{rotating}
\usepackage{txfonts}
\usepackage{color}
\begin{document} 

\newcommand{\dd}{deg$^{2    }$}
\newcommand{\flux}{$\rm erg \, s^{-1} \, cm^{-2}$}
\newcommand{\LL}{$\lambda$}

   \title{The cosmological analysis of X-ray cluster surveys}

   \subtitle{III.  4D X-ray observable diagrams}
   
   \titlerunning{The cosmological analysis of X-ray cluster surveys. III}
   
   \authorrunning{Pierre, Valotti et al}

   \author{M. Pierre\inst{1}\fnmsep\thanks{marguerite.pierre@cea.fr}, A. Valotti\inst{1}, L. Faccioli\inst{1}, N. Clerc\inst{2, 3, 4},  R. Gastaud\inst{1}, E. Koulouridis\inst{1}, 
          \and
          F. Pacaud\inst{5}
          }

   \institute{IRFU, CEA, Universit\'e Paris-Saclay, F-91191 Gif-sur-Yvette, France\\ Universit\'e Paris Diderot, AIM, Sorbonne Paris Cit\'e, CEA, CNRS, F-91191 Gif-sur-Yvette, France
         \and        
Max Planck Institut f\"ur Extraterrestrische Physik, Giessenbachstrasse 1, 85748 Garching bei M\"unchen, Germany 
\and
 CNRS, IRAP; 9 Av. colonel Roche,  F-31028 Toulouse cedex 4, France
 \and
Université de Toulouse, UPS-OMP; IRAP; Toulouse, France
\and
Argelander Institut f\"ur Astronomie, Universit\"at Bonn, 53121 Bonn, Germany            
             }

   \date{Accepted for publication in Astronomy and Astrophysics}

 
  \abstract
{Despite compelling theoretical arguments, the use of clusters as cosmological probes is, in practice, frequently questioned because of the many uncertainties surrounding cluster-mass estimates.  }
   {
   Our aim is to develop a fully self-consistent cosmological approach of X-ray cluster surveys, exclusively based on observable quantities rather than masses. This procedure is justified given the possibility to directly derive the cluster properties via ab initio modelling, either analytically or by using hydrodynamical simulations. In this third paper, we evaluate the method on cluster toy-catalogues.}
   {We model the population of detected clusters in the {\sl count-rate -- hardness-ratio -- angular size -- redshift} space and compare the corresponding four-dimensional diagram with theoretical predictions. The best cosmology+physics parameter configuration is determined  using a simple minimisation procedure; errors on the parameters are estimated by averaging the results from ten independent survey realisations.  The method allows a simultaneous fit of the cosmological parameters of the cluster evolutionary physics and of the selection effects. }
   {When using information from the X-ray survey alone plus redshifts, this approach is shown to  be as accurate as the  modelling of the mass function for the cosmological parameters and to perform better for the cluster physics,  for a similar level of assumptions on the scaling relations.  It enables the identification of degenerate combinations of parameter values.}
   {Given the considerably shorter computer times involved for running the minimisation procedure in the observed parameter space, this method appears to clearly outperform traditional mass-based approaches when X-ray survey data alone are available.  }


   \maketitle

%

\section{Introduction}

\subsection{General context}
In theory, clusters of galaxies are ideal objects for probing cosmological models because they are sensitive both to the growth of structure and to the geometry of the universe. This  includes not only constraints on the parameters of the currently favoured $\Lambda$ Cold Dark Matter paradigm ($\Lambda$CDM)  but, in principle, may probe alternative theories of gravity as well \citep[e.g.][]{brax15, cataneo15}. The basic tools rely on  cluster number counts as a function of redshift and mass, $ dn/dz, ~dn/dM/dz$ \citep[e.g.][]{planck2015XXIV, pacaud16, dehaan16}, or on topological quantities such as the correlation function 
 \citep[][]{pierre11, pillepich12} , or power spectrum, and baryon acoustic oscillations \citep[e.g.][]{balaguera11,veropalumbo16}.
However, the practical viability of such an approach to cosmology has been, and still is, very much questioned given the difficulty to properly measure cluster masses. Either because the procedures require considerable amounts of observing time to reach the necessary accuracy (for instance, with velocity dispersions or weak lensing measurements)  or because intrinsic methodological shortcomings impinge on the mass determinations. Among those, we may cite very general issues like defining cluster samples and cluster boundaries or assessing the cluster dynamical state \citep[e.g.][]{nelson14, rozo15, appelgate16}.\\
In this series of papers, we revisit the analysis of cosmological cluster surveys in the X-ray band. The X-ray  approach is of special interest since cluster properties can be routinely predicted from ab-initio modelling and considerable experience has been accumulated with the ROSAT and XMM surveys. Accurate (or supposedly accurate) masses have been determined for a number of mostly bright objects, usually assuming hydrostatic equilibrium. Subsequently, masses of fainter objects are derived from so-called scaling relations (SR), linking one observable parameter (Lx, Tx, Yx) to the mass and calibrated using the bright samples. In addition, further issues arise, which are specific to this waveband; for instance: systematic biases in the hydrostatic masses or whether or not to include cluster cores in the SR analyses. Furthermore, the question on how to model the evolution of the intra-cluster medium as a function of cosmic time and cluster mass is of special relevance: this is what determines cluster key parameters (luminosity, size, scatter in the scaling relations) that directly impact cluster detection \cite[for a review, see][]{allen11}.\\
The critical prerequisite to any cosmological analysis is the determination of the cluster selection function; in the X-ray waveband this is achieved by computing the probability of detecting a cluster as a function of flux and apparent size \citep{pacaud06}. Converting these observables into cluster masses, however, requires the knowledge of the SR. Since SR  are intrinsically cosmology dependent, in the end (i) cosmology, (ii) evolutionary cluster physics (modelled through the SR) and (iii) selection effects must be simultaneously worked out. This is a very important point, that usually requires very large computing times when following, what we name as the ``traditional bottom-up route''.  In this approach, a model is used to statistically compare predictions with data for all observed properties in both the survey data (e.g. $F_{x}, ~z$ for an X-ray survey) and follow-up measurements (e.g. $M_{gas},~ T_{x}, ~Y_{x}$, weak lensing shear) for each trial cosmology considered. Appropriate scaling relations are employed to statistically link the observed properties with mass, and each other, self-consistently accounting for the effects of survey biases and covariances 
\citep[e.g.][]{vikhlinin09,mantz10a}.  Hereafter, we subsume the various steps of this method under the dn/dM/dz or N(M,z) denomination.
In this context,  we have proposed a so-called ``top-down'' approach that consists in fitting the predicted distributions of cluster properties, such as X-ray count-rates and colours, to the observed ones \citep[hereafter paper I]{clerc12a}.

\subsection{Purpose of the present paper}
The top-down method presented in paper I was evaluated by  means of a Fisher analysis.  Considering two X-ray  observables (countrate, i.e. the number of photons collected on the detector per second in a given X-ray band - hardness ratio, the ratio of count-rates in two bands), it was shown to be more efficient than the $dn/dz$ distribution. When adding the redshift dimension, it was shown to be as good as $dn/dM/dz$ for cosmology and slightly better for the cluster physical properties. In this first validation study, cluster masses used for the comparison were assumed to be obtained from the X-ray survey data alone; this provided us with estimates for the mass uncertainties, in order to set a fair comparison between the two approaches. \\
In the present paper, we extend the method - that we call ASpiX - by adding a forth dimension, namely the cluster apparent size. This time, we evaluate the method by means of toy cluster catalogues and compare again its performance with the $dn/dM/dz$ statistics. Throughout the process, we still assume that the {\em only} source of information on cluster properties comes from a 10ks XMM large area survey. This corresponds to a medium depth of $\sim 10^{-14}$ \flux\ in the [0.5-2] keV band and yields on average some 200 X-ray photons per cluster. At this sensitivity level, reliable temperature estimates can be obtained for only about a quarter of the detected cluster population, and morphological analyses, in the form of a simple two-parameter fit, are also very much limited. Assuming such a survey allows us to ascribe realistic errors on  simple observable parameters such as X-ray flux, colours, and angular size, as well as on the cluster masses for the comparative study. We consider two survey areas of 10,000 and 100 \dd . We also investigate the effect of scatter in the mass-temperature, mass-luminosity and mass-size relation on the performances of the method.\\ 
Given this input framework, we probe the shape of the likelihood hyper-surface defined by our four-dimensional (4D) parameter space: X-ray count-rate, hardness ratio and angular size plus redshift. Practically, this is achieved by running the {\sc  Amoeba} routine on the toy cluster catalogue realisations for different sets of cosmological and cluster physics parameters. We investigate the steepness of the likelihood around the fiducial model values and the possible existence of local minima, which may reveal a degeneracy between cosmology and cluster physics, in whatever space: either the 4D observed space, or the two-dimensional (2D) M-z one. We present an empirically motivated approach to estimating the errors on the fitted parameters. \\
The article is organised as follows: Sect. 2 describes the cosmological and physical models used in the present study. Sect. 3 gives the details of the catalogue production. In Sect. 4 we describe the methodology adopted for scanning the likelihood hyper-surfaces. Section 5 presents the results of the evaluation. In Sect. 6, we discuss the results along with  a number of issues raised or left open by the current study. Section 7 gathers the conclusions and opens interesting perspectives for the use of ASpiX in the future.

\section{Description of the  cluster model}
The modelling of the cluster population, of the  X-ray properties, and of the corresponding XMM observables is largely inspired from the work of  Paper I.\\
We first compute the cluster number counts as a function of mass and redshift for a given cosmology (Sect. \ref{modecosmo}), convert masses into physical parameters (temperature, luminosity and size: Sect. \ref{modephys}), then into XMM observables (count-rate, hardness ratio, angular size: Sect. \ref{modexmm}). We apply the X-ray survey selection function and build the corresponding multi-parameter observable  diagram (Sect. \ref{modecdg}). We outline the modifications (simplification or refinements) that we have implemented with respect to Paper I, for the purpose of testing the basic behaviours of the ASpiX method.

\subsection{Cosmology and the cluster-mass function}
\label{modecosmo}
The values adopted for the cosmological parameters are those of the {\em Aardvark} simulations \citep{farahi16} - for consistency with a parallel on-going work on these simulations - namely: 
$\Omega_{m}=0.23, ~ \Omega_{\Lambda}=0.77, ~ \Omega_{b}=0.047, ~ \sigma_{8}=0.83, ~ h=0.73, ~n_{s}=1.0$. 
We use the \citet{tinker08} fit to obtain the sky-projected, redshift-dependent mass function $dn/d\Omega/dM_{200}/dz$. The equation of state of Dark Energy (DE) is assumed to depend on a single parameter $w_{0} = P/\rho$, whose value in the case of a cosmological constant is -1.
In Sect. \ref{DEdisc}, where we make a short excursion about the  case of an evolving DE equation of state, we use the parametrisation of \citet{chevallier01}: $w(z) = w_{0} + w_{a}\times z/(1 + z)$.
The parameters relevant for the present study are summarised in Table \ref{paramset}.

\subsection{Cluster physical parameters}
\label{modephys}
In paper I,  we made use of mass-temperature and temperature-luminosity relations (best determined relations for the mass range of interest). Here, we chose to scale cluster temperature, luminosity and size directly after mass. Our set of scaling relations thus reads:
\begin{equation}
\frac{M_{200}}{10^{14} M_\odot/h} = \frac{h}{0.7}\times A_{MT} \times\left( \frac{T}{4}\right)^{\alpha_{MT}}  \times E(z)^{-1}(1+z)^{\gamma_{MT}}
\end{equation}
\begin{equation}
\frac{M_{200}}{10^{14}M_\odot/h} = \frac{h}{0.7}\times A_{ML} \times \left( \frac{L}{10^{44}}\right) ^{\alpha_{ML}} \times E(z)^{-1.5}(1+z)^{\gamma_{ML}} 
\end{equation}
\begin{equation}
M_{500} \propto \frac{4}{3}\pi R_{500}^{3} ~~~ X_{c} = \frac{R_{c}}{R_{500}},  
\end{equation}

where $T$ is expressed in keV, $M$ in solar masses, $L$ in $ erg \, s^{-1} $ in the [0.5-2] keV band, $R_{c}$ is the core radius of the X-ray surface-brightness profile assuming a $\beta = 2/3$ profile, $R_{c}$ is the physical value (in Mpc) ,and we note $r_{c}$, the corresponding apparent angular size (in $arcsec$). We adopt the following conversion factor between $M_{500} $ and $M_{200}$: $M_{500} = 0.714 \times M_{200}$ \citep{lin03}. 

For each relation, we allow for scatter and assume that the three scatter values are uncorrelated. They are implemented following a log-normal distribution.  
The numerical values for the slope and normalisation of the M-T relation is from \cite{arnaud05}. For the M-L relation, they are derived from the combination of the M-T relation and the L-T relation of \cite{pratt09}. 
We consider ad hoc plausible amplitudes for the scatters and $X_{c}$, as suggested by hydrodynamical simulations \citep[as can be inferred from cosmo-OWLS,][- paper IV, in prep]{lebrun14}.
Table \ref{clusterphysics} summarises the values of the cluster physical parameters adopted for the present study.

\begin{table}
\centering                          
\begin{tabular}{l c c c c }        
\hline\hline                 
Relation & Slope & Normalisation & Scatter & Evolution  \\
$M-T$ & $\alpha_{MT}$ = 1.49 & $A_{MT} = 10^{0.46}$ & 20\%  & $\gamma_{MT} = 0*$ \\
$M-L$ & $\alpha_{ML}$ = 0.52 & $A_{ML}= 10^{0.25}$ & 50\%  & $\gamma_{ML} = 0*$ \\
$M-R_{c}$ & 3 & $X_{c}$ = 0.24* & 50\%  & none \\
\hline
\end{tabular}
\caption{{\bf Numerical values adopted for the cluster scaling relations}; (references in the text).
\newline
 An evolutionary parameter $\gamma = 0$ implements the self-similar evolution hypothesis.  
The * indicates parameters that are possibly allowed to remain free during the cosmological analysis.
} 
\label{clusterphysics} 
\end{table}

\subsection{X-ray parameters and cluster selection}
\label{modexmm}
\begin{figure}
   \centering
   \includegraphics[width=8cm]{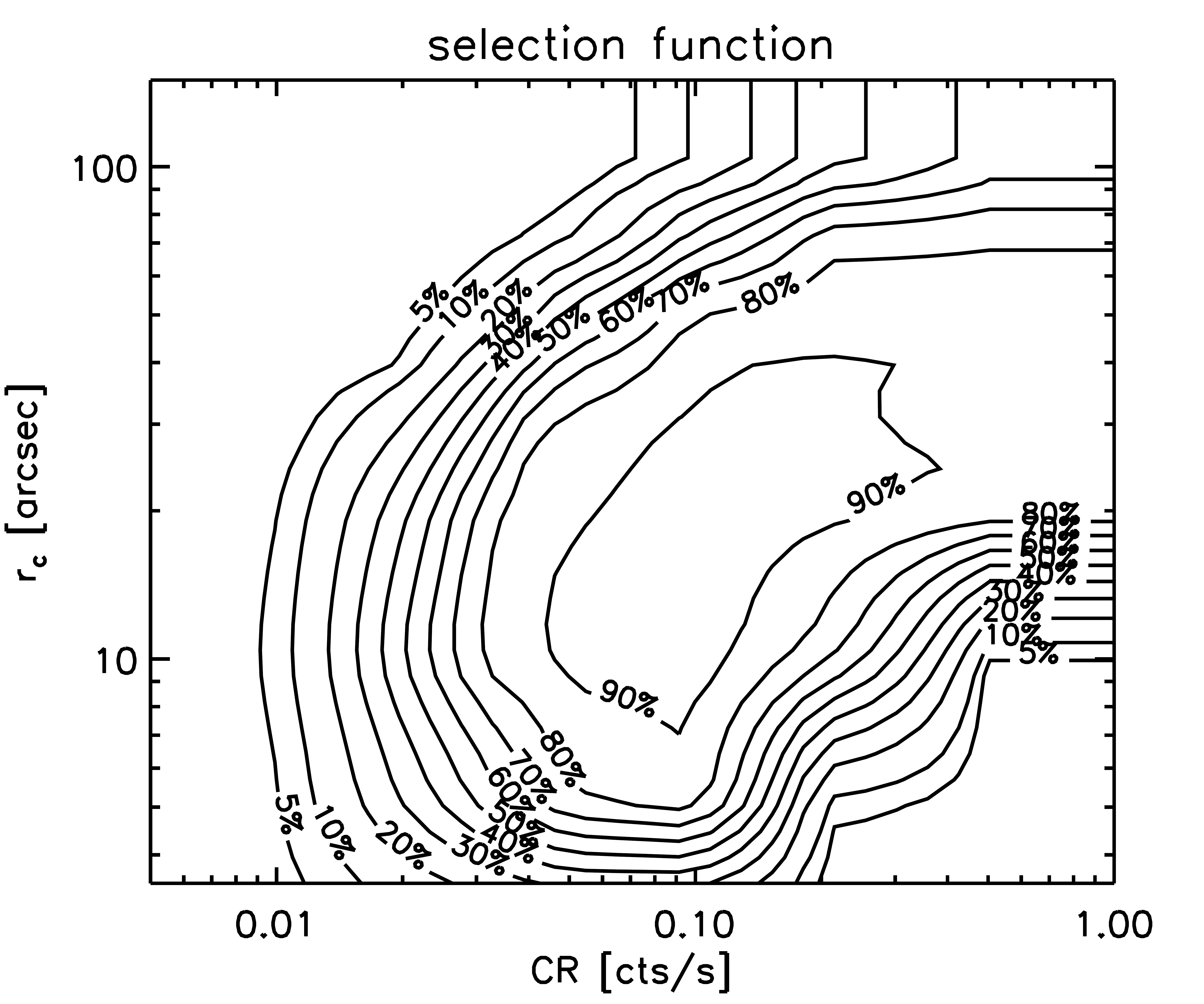}
   \vspace{0.5cm}
    \includegraphics[width=8cm]{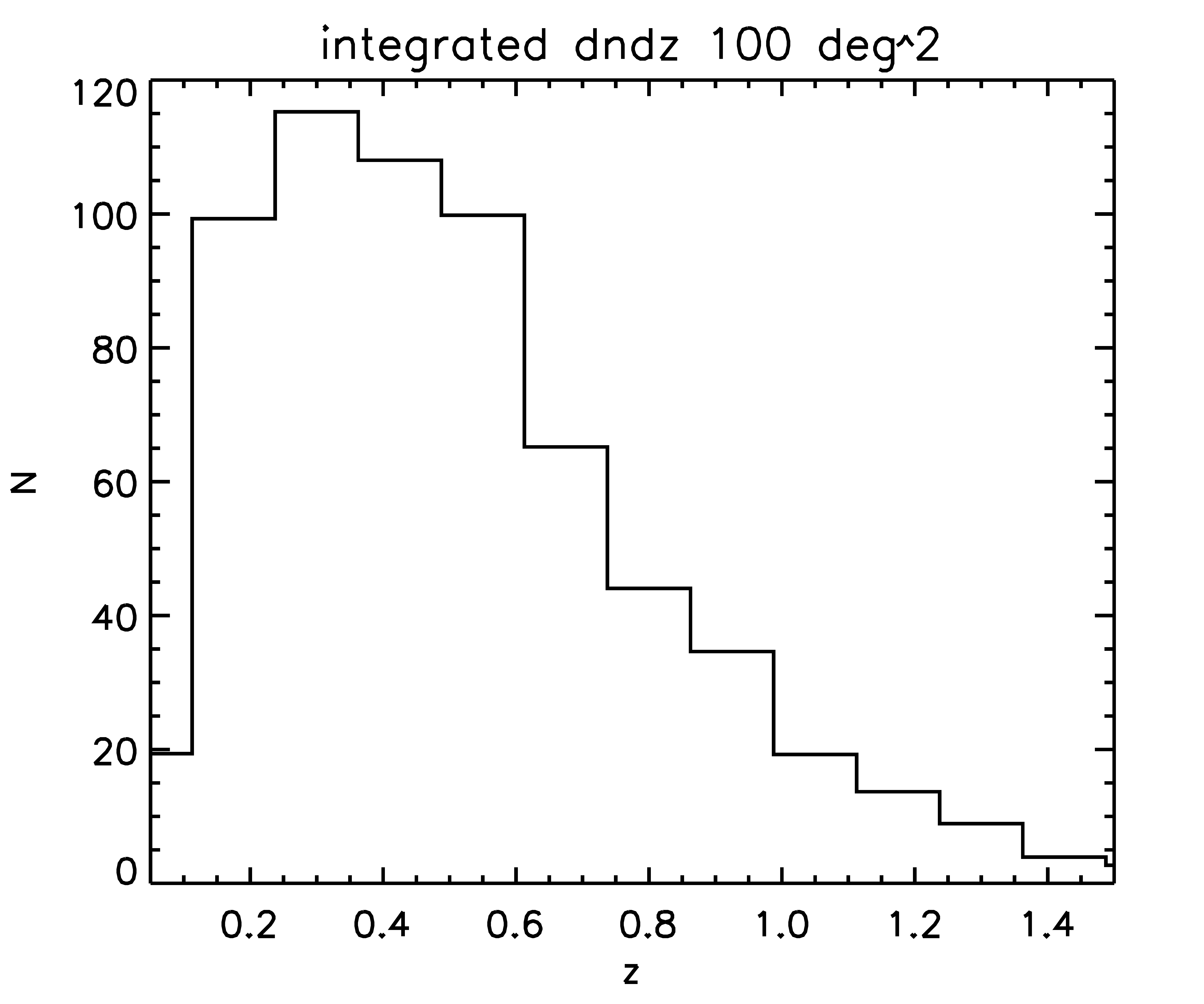} 
   \caption{Top: Selection function adopted for the present study. The probability of detecting a cluster as C1 is given by the isocontours as a function of count-rate and core-radius. This map has been derived from extensive XMM image simulations and the two axes represent the true (input) cluster parameters; it is thus only valid for the conditions (exposure time and background) under which the simulations were run.
\newline
Bottom: Resulting cluster redshift distribution for 100 \dd\ assuming the set of physical and cosmological parameters given in Tables \ref{clusterphysics} and \ref{paramset}; the cluster density is of the order of 6 /\dd .  }
              \label{selfunc}
    \end{figure}

In all that follows, we assume as in Paper I, that the X-ray clusters are extracted  from a survey, uniformly covered with 10ks XMM observations, where only sources within the inner $10'$ of the field of view are used (50\% vignetting limit in the soft band). The diffuse background is set to a uniform level of  $5.1 \times 10^{-6}$ ct/s/pixel (on-axis) plus $2.6 \times 10^{-6}$  for the soft protons  \citep{snowden08}; the unvignetted particle background is set to $2.4 \times 10^{-6}$ (Sauvageot, private communication). We model the point-source distribution, which impacts on the cluster selection function, by the flux distribution of Active Galactic Nuclei (AGN $logN-logS$) \citep{moretti03}; we assume that the AGN are randomly distributed over the XMM field of view. All this for the  [0.5-2] keV band. \\
The XMM cluster count-rates are derived from the X-ray luminosity  (which is directly obtained from the M-L relation) and redshift using the APEC code, whose output is folded with the XMM-EPIC response. 
We assume that cluster detection is performed in the [0.5-2] keV band by the {\sc Xamin} pipeline \citep{pacaud06}. Clusters are subsequently selected in the 2D {\tt Extent-Extent\_likelihood} output parameter space  of {\sc Xamin}, following the C1 criteria (Paper I). The C1 selection was shown to have less than 5\% contamination by point-sources and is calibrated using extensive XMM image simulations. In these simulations, all clusters are assumed to be spherically symmetric and to follow a $\beta=2/3$ surface brightness profile.  The resulting selection function gives the probability of detecting a cluster as a function of its  apparent size, $r_{c}$ and of its count-rate.
 Compared to Paper I, the modelling of the detection probability for  apparent sizes smaller than $10''$ has been refined by simulating clusters down to $3''$ core radius. While very few clusters are expected with such a small angular size, some of them may however be detected as C1 because of the error measurements (hence boosted above the $r_{c}>5''$ limit). The selection function and the resulting cluster redshift distribution are displayed in Fig. \ref{selfunc}.

\begin{table}
\centering                          
\begin{tabular}{l c c }        
\hline\hline                 
Name & Symbol & Unit  \\
\hline
XMM count-rate in [0.5-2] keV & $CR$ &  counts/s \\
XMM hardness ratio $\frac{count-rate ~ in [1-2]keV}{count-rate ~ in ~ [0.5-1]keV}$ & $HR$ & -\\
Apparent core radius (*)& $r_{c}$ & arcsec\\
Redshift & $z$ & - \\
\hline
\end{tabular}
\caption{Observable cluster parameters used in the cosmological diagnostic diagrams. Count-rates are corrected for the vignetting and integrated to infinity; (*) this is relative to a $\beta$-model and is the measure after deconvolution from the XMM PSF. 
} 
\label{observables} 
\end{table}

\subsection{Construction of the cluster observable diagrams}
\label{modecdg}
From the selected cluster population, we construct multi-dimensional observable diagrams (XOD) based on the cluster X-ray observable quantities at our disposal plus redshift. The main point of the present study is that these quantities are obtained from direct measurements, hence are cosmology-independent (but instrument-dependent) ; they are summarised in Table \ref{observables}. One significant upgrade compared to Paper I is the introduction of a third X-ray quantity, namely the cluster angular size, $r_{c}$, as measured on the surface brightness profile. This parameter is a potentially powerful input to the cosmological analysis as it depends both on the geometry of the space-time and on the structure growth (it is also affected by non-gravitational physics effects such as cooling and AGN feedback). Moreover, it is one of the two parameters that intervene in the cluster selection ($r_{c} \equiv$ {\tt Extent}). \\
The diagrams can be computed including, or not, scatter in the scaling relations.  The last step is to implement error measurements on the three X-ray observables. This is achieved by smearing with a log-normal distribution the relevant components of the XOD. In practice, we assume error amplitudes of $\pm 20\%$ for the three X-ray quantities (CR, HR, $r_{c}$), that are applied to all clusters, irrespective of their actual fluxes; the sigma of the log-normal filter therefore reads: 
$ \sigma = \sqrt{\phantom ~ln(0.2^{2}+1)} $.
In the present study, we assume that cluster spectroscopic redshifts are available for all clusters and have negligible uncertainties compared to the sampling used (Table \ref{binXOD}).

\subsection{Sampling the distributions}
\begin{figure}
   \centering
\includegraphics[width=8cm]{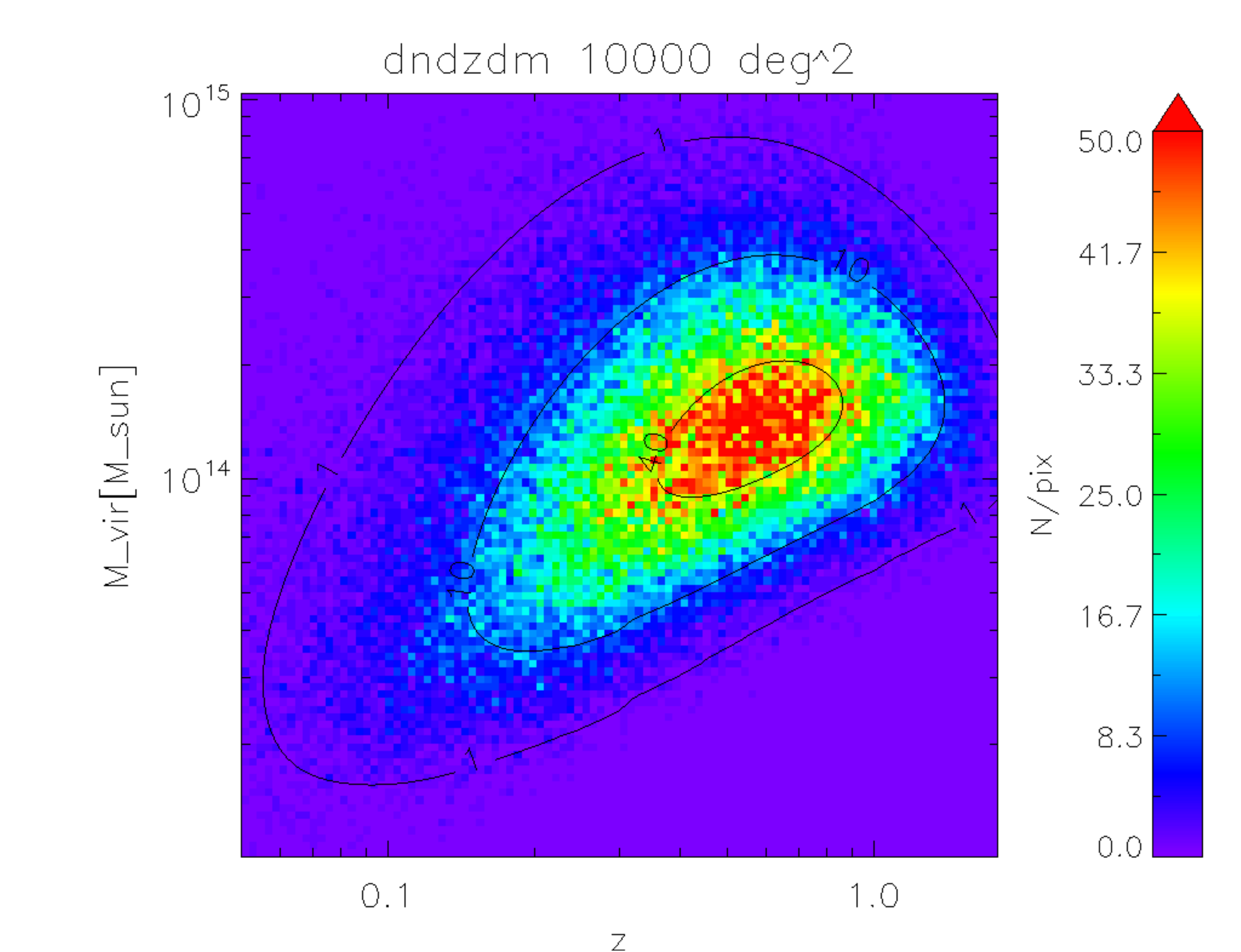}
\includegraphics[width=8cm]{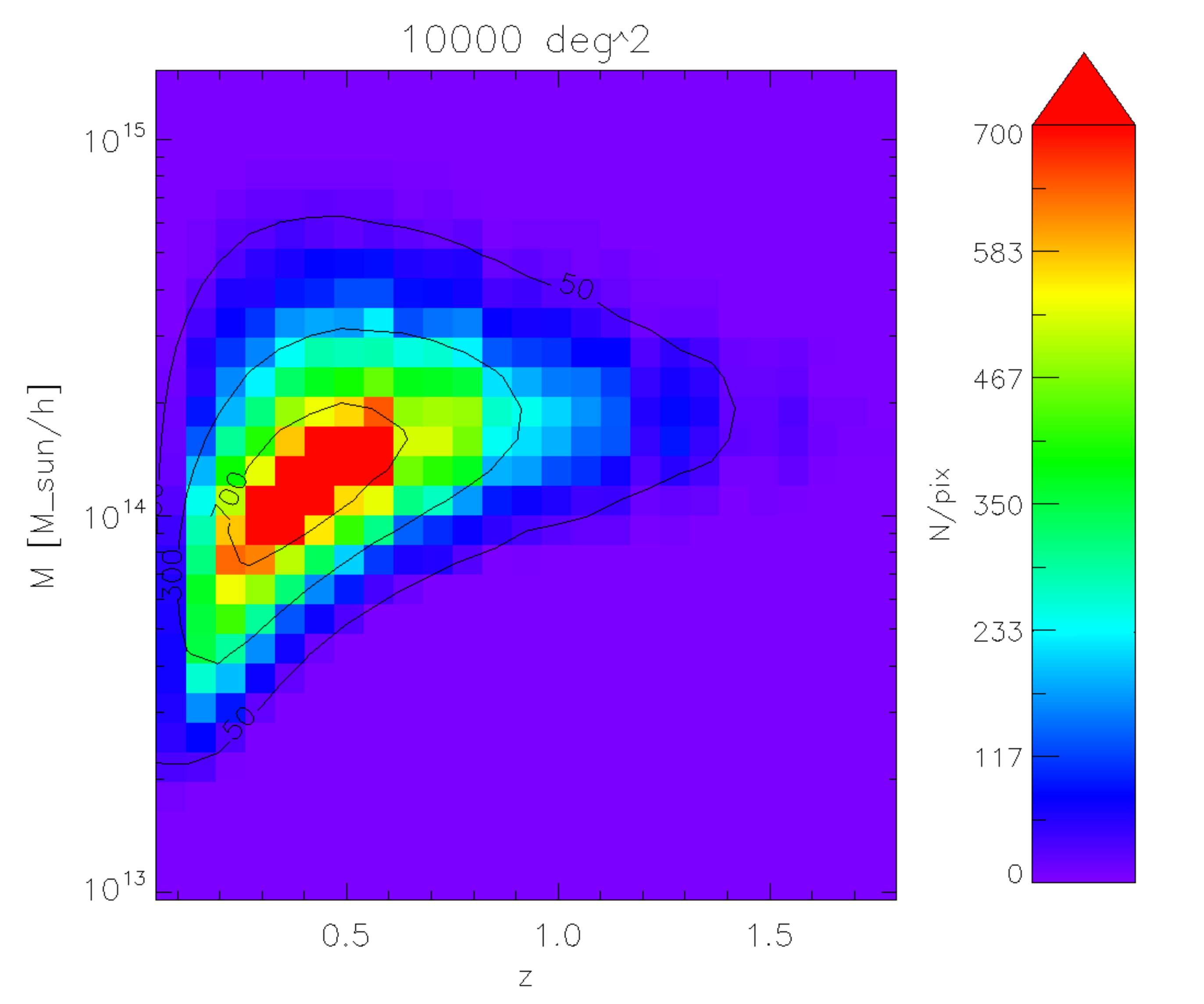}
   \caption{Top: Mass distribution  for the selected C1 population and the fiducial parameter set (including scatter in the scaling relations, which impacts the cluster selection). Cluster number densities are calculated for a 10,000 \dd\ survey. Pixels represent the catalogue (binning as in Table \ref{binXOD}) and contours are computed from the analytical model.
\newline
Bottom: same as above, rebinned  with a linear redshift scale.
      }
\label{dmdz}
 \end{figure}

\begin{figure}
   \centering
\includegraphics[width=9cm]{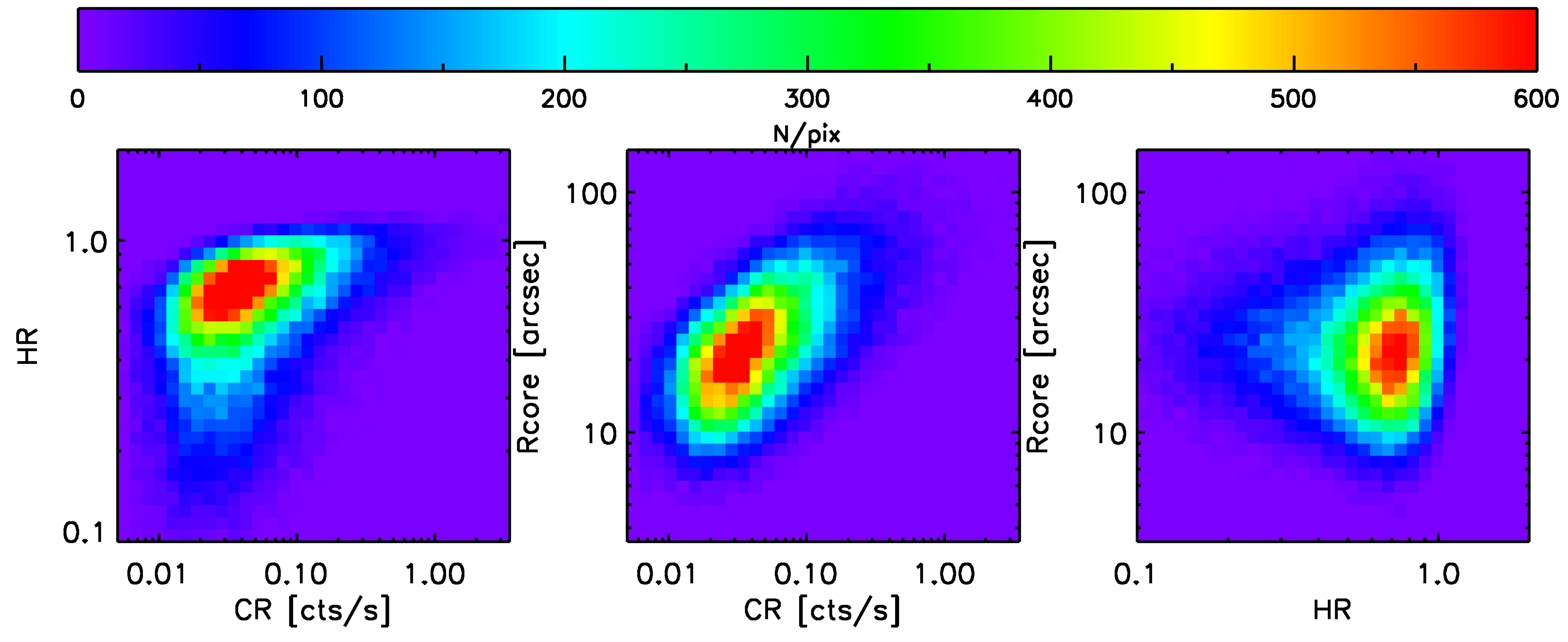}
   \caption{Projected XOD  (3D) calculated for the 10,000 \dd\ catalogue  displayed in Fig. \ref{dmdz} and integrated over the [0-1.8] redshift range. Left CR-HR; Middle CR-$r_{c}$; Right HR-$r_{c}$. The scatter in the scaling relations is implemented as in Table \ref{clusterphysics}; error measurements are not added.   }
\label{XODex1}
 \end{figure}

\begin{figure}
   \centering
\includegraphics[width=9cm]{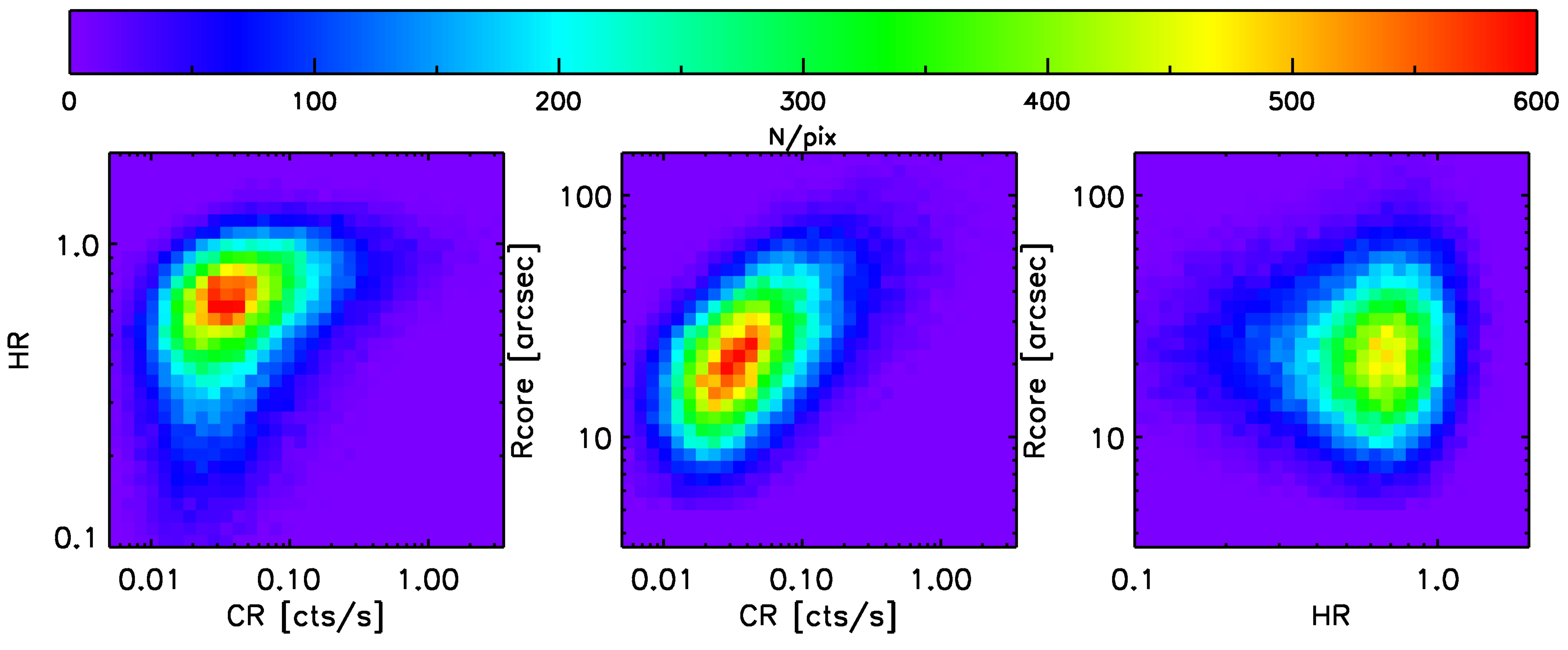}
   \caption{Same as Fig. \ref{XODex1} with a 20\% error on the measurements implemented on the three CR, HR, $r_{c}$ observables.}
              \label{XODex2}
 \end{figure}

\begin{figure}
   \centering
\includegraphics[width=9cm]{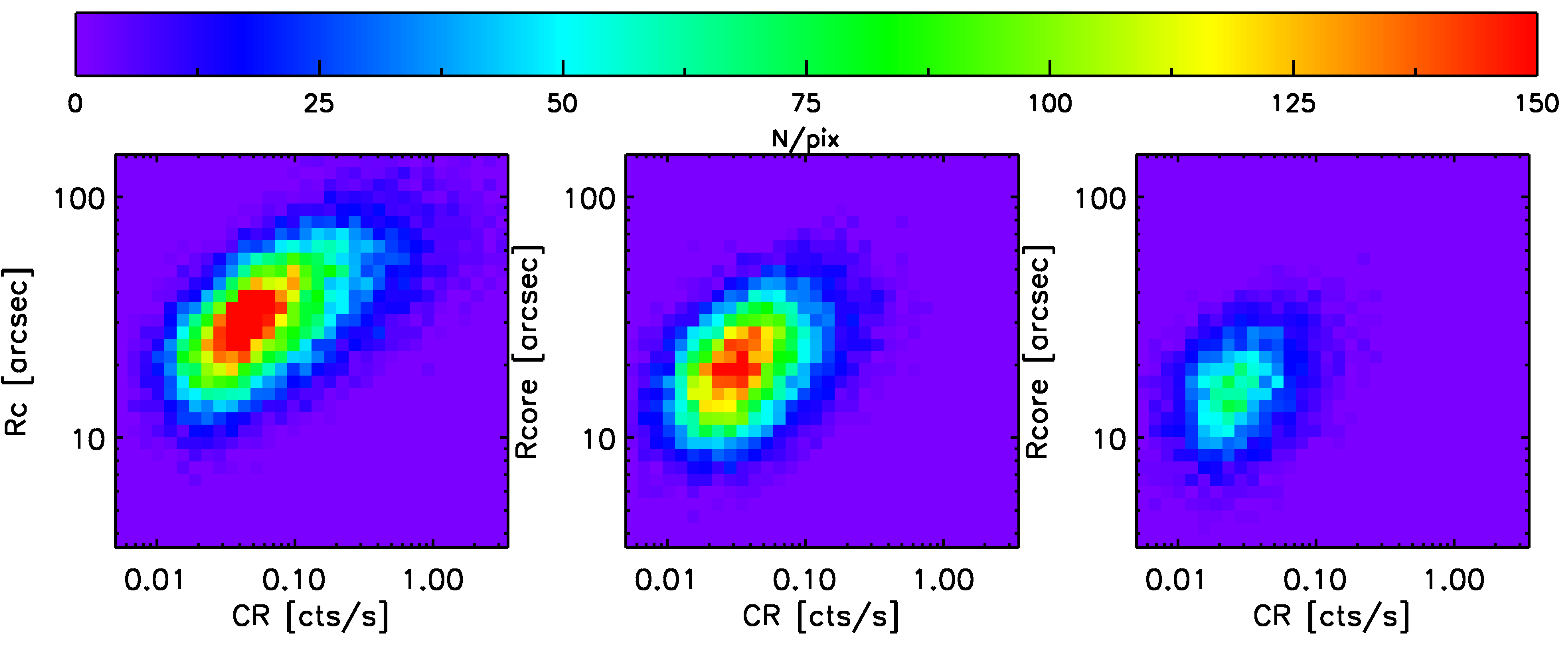}
   \caption{Projected XOD (4D) for the 10,000 \dd\ catalogue displayed in Fig. \ref{dmdz}. The CR-$r_{c}$ planes are shown for the redshift slices centred on 0.2, 0.6, 1 on the left, middle and right images, respectively. The scatter in the scaling relations is implemented as in Table \ref{clusterphysics}; the thickness of the slices is $dz=0.2$   }
\label{XODex3}
 \end{figure}

The initial cluster number counts are calculated as number densities in small dn/dM/dz bins, which, at the end of the process, are redistributed into an up to 4D observable parameter space (dn/dCR/dHR/d$r_c$/dz). In order to avoid discretisation artefacts,  it is necessary to ensure that the sampling of the input parameter space be much higher than that of the output parameter space; especially considering the steepness of the cluster mass function. To prevent overly large computing times, a trade-off in the number of bins must also be found. 
The adopted binning for the present study is given in Table \ref{binXOD}.\\
The mass distribution for the fiducial model is shown in Fig. \ref{dmdz}. Examples of corresponding projections in the XOD parameter space are shown in Figs. \ref{XODex1}, \ref{XODex2}, \ref{XODex3} and \ref{XODex4}.

\begin{table}
\centering                          
\begin{tabular}{l c c c c}        
\hline\hline                 
Parameter & Type & Min, max & Num. of bins & Scale \\
\hline
$z$ & input & 0.05, 1.8& 100 & log\\
$M$ & input& $10^{12}, ~10^{16}$ &200 & log\\
\hline
$L_{X}$ & intermediate & $10^{38}-10^{46}$ & 80 & log\\
$T_{X}$ & intermediate & 0.1-30 & 80 & log \\
$r_{c}$ & intermediate & 3.5-150 & 1024 & log \\
\hline
$CR$ & output &0.005, 3.5 & 16 & log\\
$HR$ & output & 0.1, 2 & 16 & log\\
$r_{c}$& output & 3.5, 150& 16 & log\\
$z$ (*) & output & 0, 1 & 5 & lin\\
\hline
\end{tabular}
\caption{Sampling of the input and output parameter distributions.
\newline
$r_{c}$, the apparent core radius, is used both as an intermediate parameter (intervenes in the selection function) and as an output observable (the 4th dimension of the XOD). 
\newline
 (*) : the 3D XOD are integrated over the full [0-1.8] redshift range. The 4D XOD are decomposed in 5 redshift slices over the $0<z<1$ range.
 } 
\label{binXOD} 
\end{table}

\section{Construction of the toy cluster catalogues}
In this Section, we describe the construction of cluster catalogues that will be in turn compared with the model computed in the previous Section.

\subsection{The 10,000 \dd\ catalogue.}
From the modelled mass-redshift distribution sampled as indicated in Table \ref{binXOD} and displayed in Fig. \ref{dmdz}, we compute the number of objects in each $M_{i}, z_{j}$ bin for 10,000 \dd . It is rounded up to the next integer $N_{ij}$. Then, for each $M_{i}, z_{j}$ cluster, we compute $L, ~ T, ~ R_{c}$; we scatter these three parameters following a log-normal distribution (Table \ref{clusterphysics}) and derive the corresponding count-rates and apparent core radii. The C1 selection  is applied in the $CR-r_{c}$ space (Fig. \ref{selfunc}). In this process, shot noise is modelled by applying, to each cluster, the detection probability modulated by a binomial law: retained clusters are ascribed error measurements and stored on the fly in the [CR-HR-$r_{c}$-z] XOD. The final number of clusters in this catalogue is 60700 when no scatter is implemented in the scaling relations and 63500 when scatter is taken into account.\\

From this catalogue, we infer the corresponding dn/dM/dz sample (Fig. \ref{dmdz}). This represents the best possible reference data set against which our method will be tested.

\subsection{Catalogues for 100 \dd\ }
\begin{figure}
   \centering
\includegraphics[width=9cm]{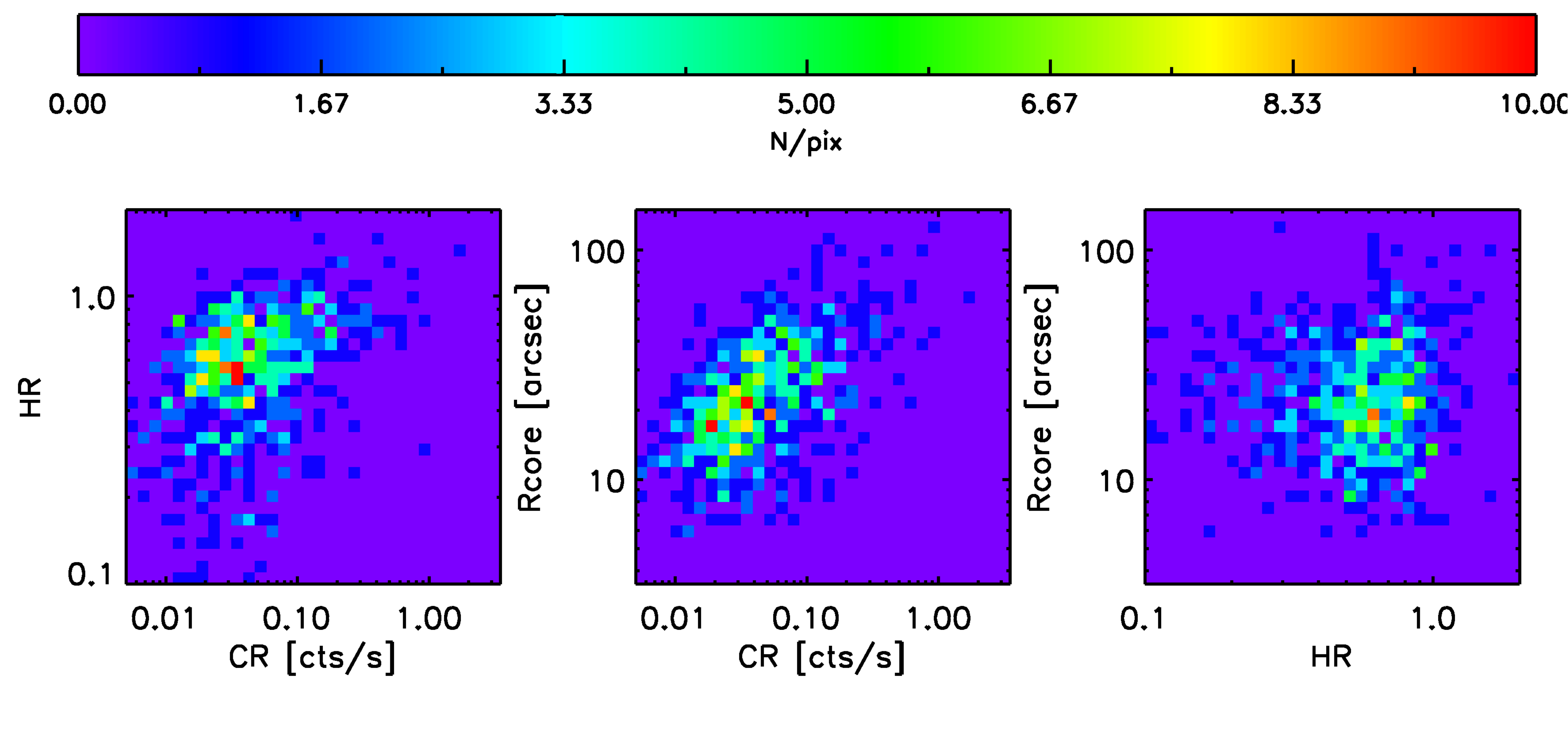}  
\includegraphics[width=9cm]{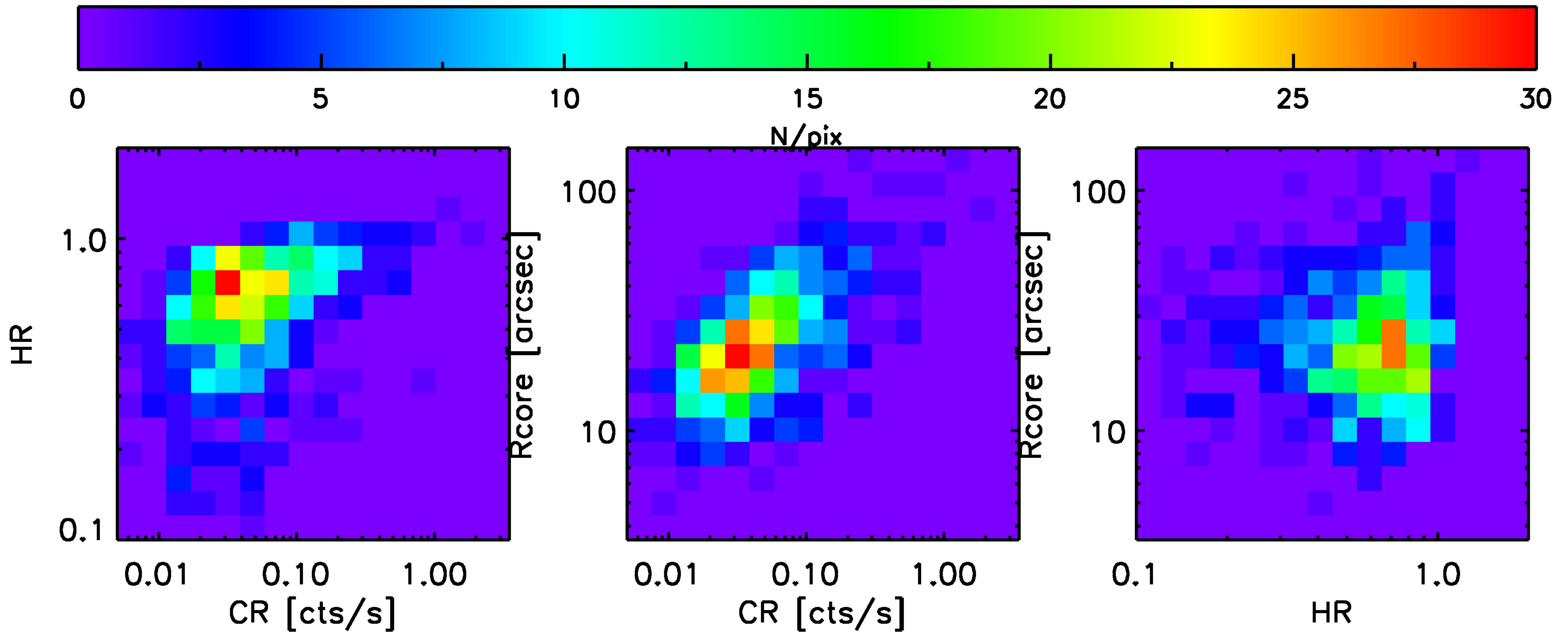}
   \caption{Same as Fig. \ref{XODex2} but for a 100 \dd\ catalogue. {\em Top:} Same binning as in Fig. \ref{XODex2}; {\em bottom:} Binning as in Table \ref{binXOD}.}.
\label{XODex4}
 \end{figure}

Catalogues for 100 \dd\ are derived from the full 10,000 \dd\ catalogue, by randomly extracting  1/100 of the objects in each redshift bin (integrated over M).  An example of a XOD diagram for a 100 \dd\ realisation is displayed in Fig. \ref{XODex4}. We have extracted ten such catalogues for each studied set of input parameters.

\begin{table}
\centering                          
\begin{tabular}{l c  } 
Parameter & Fiducial value \\       
\hline\hline                 
$\Omega_{m}$ & 0.23 \\
$\sigma_{8}$ & 0.83 \\
$X_{c}$ & 0.24 \\
$\gamma_{ML}$ & 0 \\
$\gamma_{MT}$ & 0 \\
$w_{0}$ & -1 \\
$w_{a}$  & 0\\
\hline
\end{tabular}
\caption{List of parameters on which ASpiX is tested
} 
\label{paramset} 
\end{table}

\section{Methodology for evaluating ASpiX}

We assume that the synthetic catalogues and corresponding XOD created for the fiducial model in the previous Section represent observed data sets. We describe now, how we test the ability of ASpiX to recover the initial input parameters (Table \ref{paramset}) as a function of the error measurements and of the size of the surveyed area. For this purpose, we determine the most likely solution by  scanning the input parameter space, by means of a minimisation routine. In all that follows, convergence criteria as well computer times are quoted for the IDL\footnote{http://www.harrisgeospatial.com/ProductsandSolutions/GeospatialProducts/\\
IDL/Language.aspx}
version of the codes developed for the present study.

\subsection{Likelihood estimates}

For a given catalogue, we compute the Poisson likelihood (LH) of the corresponding XOD with reference to a model:

\begin{equation}
LH = \sum_{i=1}^{N}[model_{i} -ln(model_{i})\times observed_{i}]
\end{equation}

\begin{figure}
   \centering
\includegraphics[width=9cm]{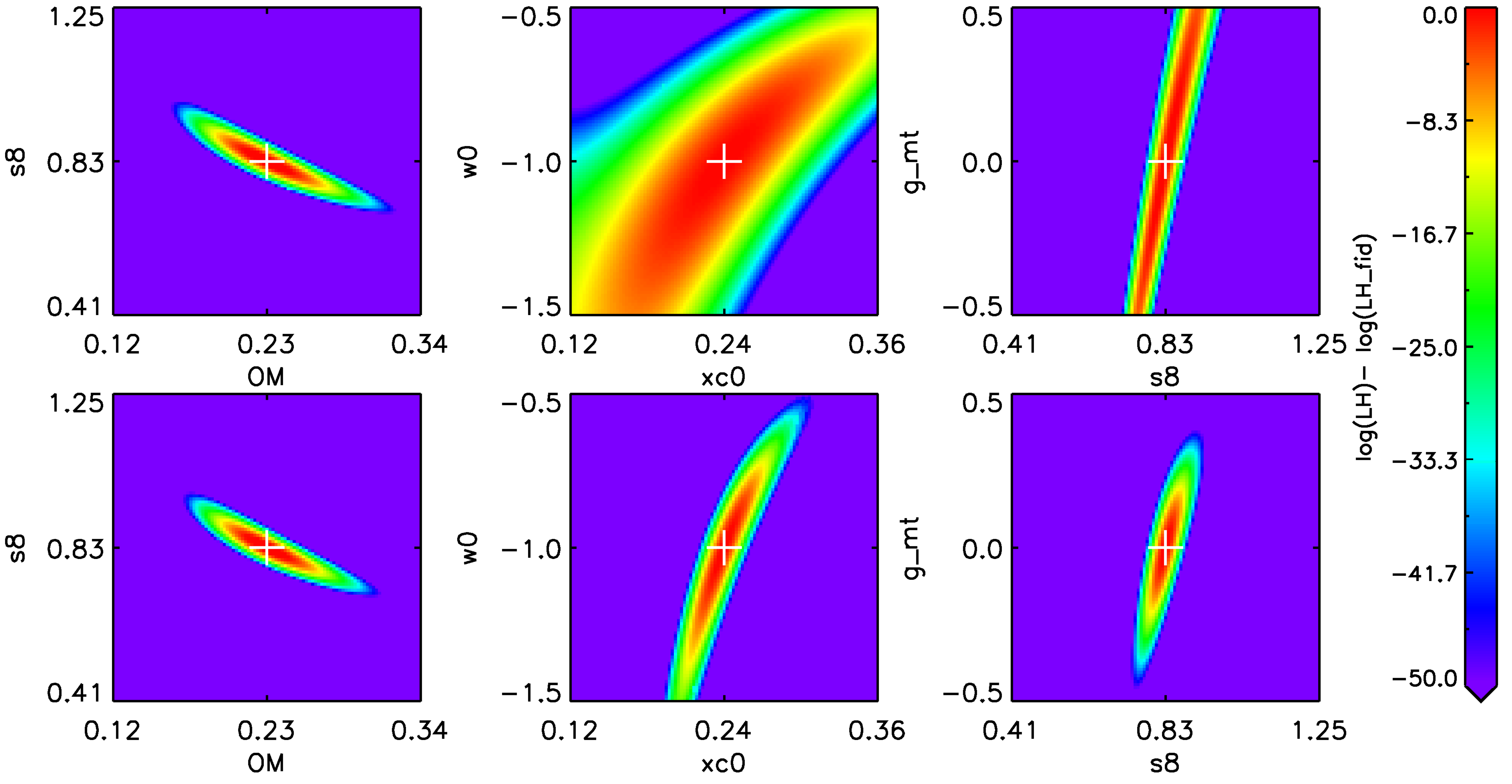}

   \caption{Top row: Slices through the likelihood hypersurface computed for a 2D CR-HR diagram.Scatter is implemented {\sl only} in the M-T and M-L relations. The likelihood surface is computed by varying only two model parameters at a time: {\em left:} $\Omega_{m}, \sigma_{8}$; {\em middle:} $X_{c}, w_{o}$; {\em right:} $\sigma_{8},\gamma_{MT}$. The central pixel corresponds to the fiducial model and has $LH=0$ (white cross). The pixel increment is 1\% of the fiducial value for all parameters. The likelihood is computed for an area of 100 \dd .
\newline   
Bottom row: same as top but this time for the 3D CR-HR-$r_{c}$ diagram. 
\newline
 }
\label{LHex1}
 \end{figure}

\begin{figure}
   \centering
\includegraphics[width=9cm]{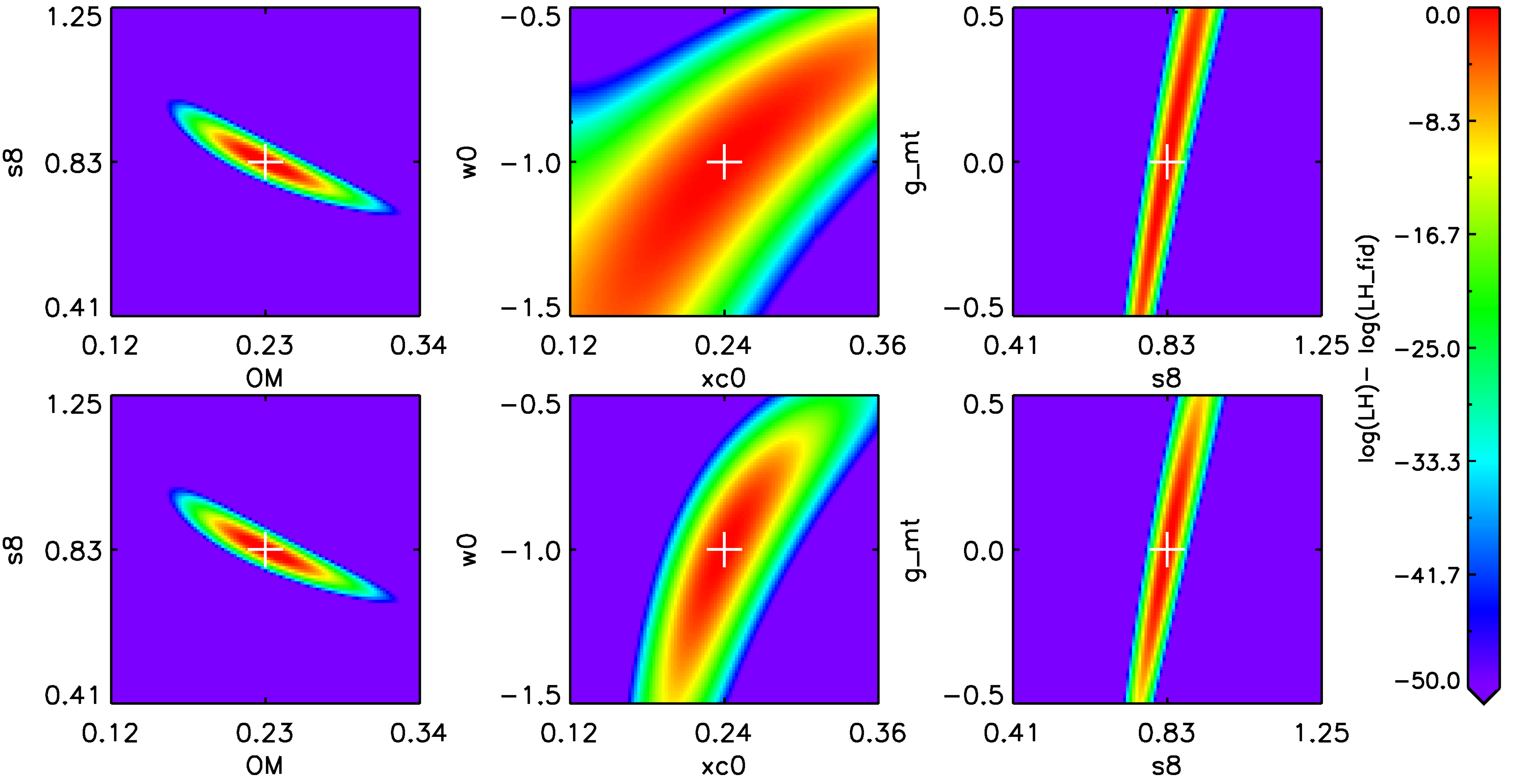}
   \caption{Same as Fig. \ref{LHex1}, but with scatter implemented in three {\sl three} scaling relations as given in Table \ref{clusterphysics}; the overall effect is the broadening of the likelihood surfaces. 
   }
\label{LHex2}
 \end{figure}

The sum is performed  over each pixel of the diagram: $model$ is the  value from the model and $observed$ is the value obtained from the catalogue. 
In order to provide a feeling for the sensitivity of the method,
we show in Figs. \ref{LHex1} and \ref{LHex2}  the LH hypersurface around the  fiducial model. This is done for a 2D and for a three-dimensional (3D) XOD.

\subsection{\sc  Amoeba}
Recovering the input parameter set from the XOD, that is, the model for which the LH is highest, presents a number of challenges: (i) flag possible degenerate solutions (i.e. different combinations of parameters that reproduce the XOD within the error tolerance), (ii) ensure that the minimisation routine does not get stuck in a local minimum, (iii) estimate the errors on the recovered parameters. In the present paper, we basically investigate the first two points and give empirical estimates for the third one.\\
In order to quickly explore the LH hypersurfaces of the XOD, we use the {\sc Amoeba} simplex method \citep{nelder65}. For a given catalogue and a given set of free parameters, we launch 100 parallel runs with different starting points for each free parameter and collect the LH and parameter values at the position where each run stops. As convergence criterion we set:
\begin{equation}
2 \times \frac{LH_{max}-LH_{min}}{LH_{max}+LH_{min}} < 10^{-5},
\end{equation}
where $LH_{max}$ and $LH_{min}$ are the extreme values obtained for a given  simplex (this tolerance is relaxed to $10^{-3}$ when more than five parameters are fitted at a time).
Given a set of 100 realisations, we construct the following statistics for each fitted parameter: histogram of the end-values, end-values versus LH, end-values versus starting point. We then search for the solutions having the highest likelihood values, assuming that less significant end-points correspond to trajectories that end up in local minima. In particular, we consider, the solution corresponding to the highest LH as well as the average of the solutions given by the 10 and 20 highest LH trajectories namely: best, best-10, best-20.  

\section{Evaluation of ASpiX}

\subsection{Test configurations}
\label{testrun}

We first set a benchmark by assuming an ideally perfect (and perfectly utopian) cluster survey. For this, we assume that the N(M,z) distribution for the detected clusters is totally under control: there is no error on the mass measurements and selection effects as a function of mass and cosmology are fully monitored. In a subsequent step, we ascribed a 50\% mass uncertainty to all detected clusters.  These will be the references for testing the ASpiX performances (runs M1 to M5). \\
In  the remainder of this paper, we mostly explore the behaviour of ASpiX with four free parameters. We use part of,  or the full observable space: CR, HR, $r_c$, z; that is, from 2D (as in paper I) to 4D XOD.
We investigate the impact of shot noise by considering, in addition to an area of 10,000 \dd ,  ten or more realisations of 100 \dd\ catalogues. The range of starting values injected in {\sc {\sc {\sc Amoeba}}} is $\pm 50 \%$ of the fiducial values of all parameters.\\
The list of tested configurations is given in Table \ref{runlist}.

\begin{sidewaystable*}
\centering                          
\begin{tabular}{c c c c c  ||c c c c c c}        
\hline\hline                 
Run ID ~~(1)& Diagrams ~~(2) &   Area ~~(3)& S ~~(4)& E ~~(5)&$\Omega_{m}$ ~~(6)& $\sigma_{8}$ ~~(7)& $X_{c}$ ~~(8)& $w_{0}$~~(9)&$\gamma_{ML}$~~(10)& $\gamma_{MT}$ ~~(11)\\
\hline \hline
M1 & M, z &    10,000 &  & N &0.23 &0.83 & 0.24& -1.00& -&- \\
M2 & M, z &   [$10 \times $] 100 & & N & 0.23 $\pm$ 0.010& 0.82$\pm$ 0.010 & 0.24$\pm$ 0.015& -1.01 $\pm$0.070&  - & -\\
M3 & M (50\% err), z &  [$10 \times $]  100 &  & Y& 0.24 $\pm$ 0.015& 0.81 $\pm$ 0.015 & 0.24 $\pm$ 0.025& -0.99 $\pm$ 0.10& -& - \\
M4 & M (50\% err), z  &  10,000 &  &Y & 0.23& 0.82&0.24 &- &-0.03 & -0.11 \\
M5 & M (50\% err), z &  [$10 \times $]  100 &  & Y& 0.23 $\pm$ 0.025& 0.82 $\pm$ 0.015& 0.24 $\pm$ 0.010& -& -0.12 $\pm$ 0.25 & -0.40 $\pm$ 1.60 \\
\hline
A0 & CR, HR &  10,000& N & N & 0.23& 0.83  &0.24 &-1.00 &- & -\\
A1 & CR, HR &  10,000& Y & N & 0.23& 0.83  &0.23 &-1.06 &- & -\\
A2 & CR, HR, $r_{c}$& 10,000& Y & N & 0.23&  0.83&  0.24 &-1.06 &- & -\\
A3 & CR, HR, $r_{c}$, z & 10,000& Y & N &0.23 &0.83 & 0.24  &-1.03  &- & -\\
A4 & CR, HR, $r_{c}$, z &  10,000& Y & Y  &0.23 &0.83   & 0.24&-1.04 &- & -\\
\hline 
B1 & CR, HR & [$10 \times $] 100 & Y & N  & 0.23 $\pm$ 0.040 & 0.82 $\pm$ 0.035 & 0.24 $\pm$ 0.030& -1.23 $\pm$ 0.55&-& -\\
B2 & CR, HR, $r_{c}$ & [$10 \times $] 100 & Y & N  & 0.23 $\pm$ 0.025 & 0.83 $\pm$ 0.025 & 0.24 $\pm$ 0.020& -1.05 $\pm$ 0.40&- & -\\
B3 & CR, HR, $r_{c}$, z & [$10 \times $] 100 & Y & N  & 0.23 $\pm$ 0.015&0.83 $\pm$ 0.015  & 0.24 $\pm$ 0.006& -1.01 $\pm$ 0.090&- & -\\
B4 & CR, HR, $r_{c}$, z &  [$10 \times $] 100 & Y & Y  & 0.23 $\pm$ 0.010& 0.82 $\pm$ 0.015& 0.24 $\pm$ 0.005&-1.00 $\pm$ 0.090 &- & -\\
\hline 
A6 & CR, HR, $r_{c}$, z &  10,000 & Y & Y & 0.24& 0.82& 0.25& -&0.020 &0.06 \\
B6 & CR, HR, $r_{c}$, z &  [$10 \times $] 100 & Y & Y & 0.23 $\pm$ 0.020& 0.82 $\pm$ 0.050 & 0.24 $\pm$ 0.010& -& 0.00 $\pm$ 0.20& 0.00 $\pm$ 0.25\\
\hline \hline
Model & & && &  0.230& 0.830& 0.240&  -1.00 & 0.00 &  0.00\\
\hline \hline
\end{tabular}
\caption{{\bf List of test configurations and result summary}
\newline
Column 3: in \dd\
\newline
Columns 4: whether scatter is implemented in the scaling relations. 
\newline
Column 5: whether error measurements are taken into account: 20\% log-normal on CR, HR, $r_{c}$; 50\% log-normal on the mass.
\newline
Columns 6 to 11: Results of the fits. A ``-'' indicates that the parameter was held fixed to its fiducial value. When no error is given, only one catalogue (10,000 \dd ) was fitted and the quoted value is the best-10. For the 10 $\times$ 100 \dd\ realisations, the quoted value is the mean of the ten best-10 and the  corresponding dispersion is assumed to be representative of the 1$\sigma$ accuracy (the last digit is rounded to the next half integer).
} 
\label{runlist} 
\end{sidewaystable*}

\subsection{Results}
 
The results are summarised in Table \ref{runlist} and comprehensively presented in Appendix \ref{output}.
In order to simply quantify the performances of the method, we assume that the rms of the output from at least ten different 100 \dd\ catalogues provides a sensible accuracy indicator. We shall discuss this assumption in Sect. \ref{disc-precis}. 

\subsubsection{Running {\sc Amoeba} on the mass distribution}
Configuration M1 yields, as expected, a solution that is  extremely close to the fiducial values (less than 0.4 \% offset for each parameter).  Dividing the survey area by 100 and considering the average output from ten different catalogues (M2) we find a mean error on $w_{0}$ of 7\% and less for $\Omega_{m}, ~ \sigma_{8}, ~ X_{c}$. Then, assuming 50\% error on the mass for all detected clusters (M3), the uncertainty on $w_{0}$ reaches $\sim$ 15\% and less for the three other parameters. 

\subsubsection{Running {\sc Amoeba} on the X-ray observable diagrams}
In the second part of the exercise, we test step by step, the behaviour of ASpiX over a 10,000 \dd\ area (runs A0 to A4). This single catalogue realisation  is intended to provide an example of the ultimate accuracy reachable by ASpiX. The A0 configuration (CR-HR only, no scatter in the scaling relations, no error measurements) converges towards a solution that gives  $\Omega_{m}, ~ \sigma_{8}, ~ w_{0}$ at the $10^{-3}$ accuracy level. Adding scatter (A1) leads to a 6\% offset on $w_{0}$ for this particular catalogue realisation. Adding the third dimension, $r_{c}$ has no noticeable effect (A2).  The knowledge of cluster redshifts (A3), reduces  the offset to 3\% . Assuming, finally, error measurements of 20\% for the three X-ray observables, the accuracy on $w_{0}$ recedes to 4\% but the other parameters remain basically unaffected (A4).\\ 
We now turn to the results pertaining to the 100 \dd\ catalogues, to investigate the behaviour of ASpiX for more realistic survey coverages.
The improvement when switching from B1 to B2 (adding the $r_{c}$ dimension) is more conspicuous than for  the 10,000 \dd\ realisation, as suggested by the $1\sigma$ deviations obtained by averaging the best values from the ten catalogues. Taking these as face error values, we observe an improvement by a factor of $\sim 1.5$ for $w_{0},\Omega_{m}, ~ \sigma_{8}$ . Implementing the redshift information  produces a drastic effect: the uncertainty on $w_{0}$ is divided by a factor of 5 (B3). Finally, modelling the impact of 20\% error measurements (B4) has little effect; it suggests that we can estimate $w_{0}$ at better than 10\% and reach at least a 2\%  accuracy for the other parameters.\\
In addition to the study of the [$\Omega_{m}, ~ \sigma_{8}, ~ X_{c}, ~w_{0}$] combination, we found it relevant to investigate the ASpiX ability to pinpoint cluster evolution trends. Indeed, this issue (whether self-similar, strong, weak, positive, negative evolution . . . ) has been generating considerable debate for the last decade without yet ending in a firm conclusion; at least it has been clearly established that selection effects and the form of the scaling relations assumed for the local universe play a critical role (for a discussion, see \cite{giles16}). For this purpose, we study the [$\Omega_{m}, ~ \sigma_{8}, ~  X_{c}, ~\gamma_{MT}, \gamma_{ML}$] parameter combination (configurations A6 and B6). The $1\sigma$ uncertainty as indicated by the average of 10 x 100 \dd\ runs for the 4D XOD, including scatter and error measurements, suggests that (i) we can obtain $\Omega_{m}, ~ \sigma_{8}, ~ X_{c}$ at better than 10\% and (ii) the M-T and M-L evolution parameters appear both to be within  $0 \pm 0.25$ (1$\sigma$ errors). 
 For comparison, the N(M,z) statistic (M5 configuration) shows similar accuracy for $\Omega_{m}, ~ \sigma_{8}, ~\gamma_{ML} $ but appears much poorer for  $\gamma_{MT}$. 
\\
In summary,  the overall performance of ASpiX with the 4 D XOD  appears to give  constraints comparable to the idealised N(M,z) statistic (with a much better score however  for $\gamma_{MT}$)  when assuming realistic error measurements for all parameters. From the point of view of the methodological simplicity, this is a considerable improvement.

\section{Discussion}
\label{discussion}

We discuss below the main outcome of this validation study and address a number of issues raised by the adopted methodology.

\subsection{Estimated precision of the parameters.} 
\label{disc-precis}  
Our conclusions mainly rely on the assumption that averaging the results from at least ten different catalogue realisations provides a sensible estimate of the marginalised  1$\sigma$ uncertainties. In order to test this, we ran a  Fisher analysis (FA) for the studied configurations. For instance, given the [$\Omega_{m}, ~ \sigma_{8}, ~ X_{c}, ~w_{0}$] parameter set, the FA predicts: for  M3 $ \pm 0.014, ~ \pm 0.014, ~\pm 0.020, ~\pm 0.085$ to be compared with the {\sc Amoeba}-derived estimates (0.015, 0.015, 0.025, 0.10); For B4, we obtain $\pm 0.012, ~ \pm 0.021, ~\pm 0.007, ~\pm 0.07$  to be compared with the {\sc Amoeba}-derived estimates (0.010, 0.015, 0.005, 0.09); these results are very similar. For the B6 configuration that handles the [$\Omega_{m}, ~ \sigma_{8}, ~  X_{c}, ~\gamma_{MT}, \gamma_{ML}$] parameter set, the FA predicts $ \pm 0.015, ~ \pm 0.046, ~\pm 0.010, ~\pm 0.22 ~\pm 0.19$ while
the {\sc Amoeba}-derived estimates give (0.020, 0.050, 0.010,  0.25, 0.20). 
The concordance with the FA analysis is very encouraging   and we foresee that using significantly more than ten catalogue realisations will render the error assessment even more reliable. 
The next step will be  to cross-check these error estimates using   cosmological N-body simulations of the cluster population \citep[see e.g. Sect. 5 in][]{pierre16}

\subsection{The use of {\sc  Amoeba}.}
We chose {\sc Amoeba} to evaluate the XOD performances along with a cluster toy model because it  converges quickly. Moreover it allows us, by using a wide range of starting values, to scan a large number of points of the LH-hypersurface and thus, explore possible degeneracies.   The drawback being that, compared to a full MCMC implementation, it does not provide errors on the fitted parameters. But as shown above,  reasonable error estimates can be obtained by averaging the output from different catalogue realisations. Also, from the practical point of view, it is straightforward to launch 100 independent parallel runs on a single catalogue.
Our results show that for up to four free parameters fitted simultaneously: (i) we do not observe any correlation between the starting and end values; and (ii) solutions obtained from the best LH, the averaged 10-LH, and 20-LH are similar in general.
We note that, while we restrain the starting values to $\pm 50\%$ of the fiducial solution, points much further out are explored in the course of an {\sc Amoeba} run (see e.g. Fig. \ref{M7degenerate}). The situation where the parameter set would very significantly differ from the favoured concordance model will be addressed in a future paper.\\

\subsection{A fair comparison with N(M,z) ?}
We compare the ASpiX performances with the constraints obtained from the mass distribution alone. In doing this, we assume that the mass for all clusters is estimated at the 50\% accuracy level. To ensure a fair comparison with the XOD, we consider this accuracy to be reached by using, in some way, the 10ks XMM data (only). Given, for example, that cluster temperatures can be measured for about 75\% of the entire C1 population and that, for about half of the cases, the errors are larger than 50\% \cite[][Table D1]{pacaud16}, this assumption on the mass accuracy is optimistic. 
In that sense, we are confident that assessing the performances of ASpiX with respect to the constraints from the mass distribution modelled in that way, provides safe conclusions.
Of course, masses can be more precisely evaluated by performing deeper X-ray observations or/and adding information from weak lensing, but then we undermine our challenging goal of providing a cosmological analysis with X-ray survey data alone.

\subsection{Added value from cluster size information}
A major upgrade from  paper I is the introduction of the angular core radius ($ r_{c}$) as the third X-ray observable parameter. A priori, it is expected to play a significant role because it directly enters the selection function (Fig. \ref{selfunc}) and is a  function of  angular distance, and hence of cosmology ($r_{c} \equiv R_{c}/D_{a} = R_{500} \times X_{c}/D_{a}$). 
A first impression of the added power is given by Figs. \ref{LHex1} (no scatter on $X_{c}$) and \ref{LHex2} (scatter), in which we show the impact of $r_{c}$ when no redshift information is available and only two parameters are let free, namely for the chosen examples: $\Omega_{m}-\sigma_{8},~ X_{c}-w_{0}, ~ \sigma_{8}-\gamma_{MT}$.  Without the implementation of $r_{c}$, a degeneracy exists between cluster intrinsic radii (hence masses) and cosmology, namely between $ X_{c}$ and $w_{0}$ (top central panels).  The likelihood surface is significantly sharpened with the introduction of the $r_{c}$ measurement (bottom central panels). Conversely, this appears to have no or moderate impact for the $\Omega_{m}-\sigma_{8}$ and $\sigma_{8}-\gamma_{MT}$ combinations respectively; the latter being almost washed out when the  scatter in the scaling relations is implemented. It is expected however, that the introduction of  $r_{c}$  has a greater impact when more parameters are let free. This can be appreciated by comparing configurations B1 and B2: the 1$\sigma$ error obtained on $ r_{c}$ and $w_{0}$ are divided by $\sim$ 1.5 when introducing the  $ r_{c}$ observable. The improvement for the 10,000 \dd\  (A1 and A2) is not straightforward to estimate, since we consider only one such mock catalogue - it is also possible that because of the very large number of clusters plus the fact that for these configurations we assume no measurement errors,  we have already reached the maximum information that can be extracted from the X-ray data given the convergence criterion imposed; or that the sampling used here with the 16x16x16 (CR, HR, $r_{c}$) grid is too coarse for this precision level. 
Interestingly, we note that $X_{c}$ is already well constrained by the CR-HR parameter space alone, as can be also appreciated by comparing runs B1 and B2.

\subsection{The Dark Energy equation of state ?}
\label{DEdisc}
Very large future cluster surveys ultimately aim at determining the DE equation of state in a standalone fashion or in combination with other cosmological probes \citep{albrecht06}. To explore the behaviour of ASpiX in this more general context, we have extended the analysis to a situation where all cosmological parameters but the evolving DE equation of state are determined at a high level of accuracy (e.g. by the Planck CMB analysis). In this configuration, we assume that the parameters of the scaling relations are unknown (scatter excepted), which corresponds to a set of nine free parameters in total. Results are displayed in Table \ref{runA10} along with the comparison with the N(M,z) statistics as previously (runs M10 and A10).   Here also, we estimate the uncertainty on the parameters, by averaging the results of ten different catalogues. The accuracy on $w_{0},~w_{a}$  appears to  be comparable for the XOD and N(M,z) approaches, but with a significantly lesser performance for N(M,z) on the M-T relation: this is due to the fact that the M-L relation, only, directly intervenes in the cluster selection and that the role of the temperature has not been explicitly coded in the global error budget of the cluster masses. This demonstrates the power of ASpiX to simultaneously determine cosmological and scaling relation parameters.     
 Switching to a 100 \dd\ survey and assuming that the scaling relations are known gives a 
15\% uncertainty on $w_{0}$ when $w_{a}$ is let free; this latter parameter could be constrained better than $\pm 1$. The N(M,z) statistics (run M12) shows errors some 1.5 times larger.
Another example of the power of ASpiX for the DE equation of state is illustrated in Appendix \ref{degeneracy} (see following point). We defer the detailed evaluation of ASpiX performances for the DE to a future publication.

\subsection{Degeneracy between physics and cosmology}
An interesting application of running {\sc Amoeba} with 100 different combinations of starting values is the exploration of possible degenerate sets of parameters, as rendered by the imprint of local minima in the likelihood hypersurface. This is a quick way to probe the surface, but of course it does not replace a systematic (and extremely computer-time-consuming) approach. Moreover, because of statistical fluctuations, some catalogues may favour a specific parameter configuration. Hence,  ultimately, we need be able to discriminate between a true degeneracy, that is, two different parameter sets yielding very similar XOD for 10,000 \dd\,  and large uncertainties.  We have investigated a few cases in some detail by means of the 100 \dd\ catalogues. The presence of degeneracies is suggested by a significant difference between the average of the parameter values returned by the best LH and the average of all values, when merging the output from all catalogues. We illustrate this fact by two cases, spotted in the course of the simulations (Appendix \ref{degeneracy}). Quite surprisingly, the second one shows that a degeneracy can exist in the mass distribution, while being totally resolved by the XOD 3D representation.
Our current understanding of the situation is that there is no strong degeneracy   in the 4D XOD space at the precision level given by 10 ks XMM data and 100 \dd . Conversely,  obvious ones (at least one) seem to be present in the N(M,z) space, even though this corresponds to an exotic $w_{a}-w_{0}$ combination.

\subsection{Neglected issues}
In this first systematic study, we made a number of simplifications: (1) 
We assumed that the (local) scaling relations are perfectly known and fixed (except for configuration A10, where all scaling parameters are let free); from the practical point of view, this is incorrect since luminosities as well as mass measurements, are cosmology dependent. However, for the adopted fiducial model, a 25\% increase in $\Omega_{m}$ induces a $\sim$ 2\% increase in the luminosity distance at $z=0.5$, that is, some 4\% in luminosity estimates, which we consider as negligible for the demonstration purpose of the paper, given the assumed error measurements.  
(2) Scatter in the scaling relations is also fixed and constant with redshift. This is an optimistic assumption, given that even for the local universe, scatter values are still poorly constrained  and that they play a critical role in the selection function (\cite{pacaud07}, paper IV). Moreover, we assumed that the scatters in the M-L and M-T SR are independent as well as the errors on CR, HR and $r_c$; covariance certainly plays a role here. 
(3) We neglected sample variance effects as well as the uncertainties linked to the assumptions of the Tinker mass function and of the random location of AGN as well as  other more subtle effects such as the non-universality of the halo mass function for different DE models or non-minimal neutrino masses.   We defer the implementation of more complex scaling relations than simple power-laws along with uniform Gaussian scatter to paper V, which will apply ASpiX on hydrodynamical simulations. \\
These simplifications should  similarly affect the ASpiX  and N(M,z) methods, therefore do not effect the conclusions of the present comparative study. 

\subsection{Computing speed}
As a rule of thumb, the  computation time of an XOD diagram  increases linearly with the number of observables; it is multiplied by a factor of 4 when adding scatter. The error implementation has a negligible impact  given the simple error model assumed here. Moreover, the running time for the cosmological analysis is roughly proportional to the number of fitted model parameters (Sect.  \ref{testrun}). For instance, the fitting of realisations A0 and A4 takes 22 min and 1h30, respectively, for  100 independent {\sc Amoeba} runs distributed over 100 CPUs with the IDL version of the code. For a given configuration, not all {\sc Amoeba} runs converge: the convergence rate varies from 90\% to 60\% (when more parameters are added).

\subsection{Using more X-ray bands ?}
A natural question arising at this stage, is whether or not we have exhausted all possible X-ray information. Strictly speaking, the answer is no, since we did not make use of the X-ray photons from the [2-10] keV range. In theory, a pure bremsstrahlung spectrum is uniquely determined by a single hardness ratio and the redshift-temperature dependence scales as $T \propto 1/(1+z)$ if we neglect the Gaunt factor. Emission lines certainly have an impact, especially the presence of the Fe complex at 6.7 keV restframe, that provides implicit redshift  information at our sensitivity. However, given the low number of cluster photons collected in the [0.5-2] keV band during 10ks exposures ($\sim 200$ in average), the fact that  the XMM sensitivity significantly drops beyond 2 keV and the increasing particle background level at high energy, we anticipate that adding a  fifth dimension to the diagram (e.g. the CR in the [2-5] keV band, or a new hardness ratio) will basically have little effect, if not simply add noise (paper IV).

\begin{sidewaystable*}
\hspace{-1cm}
\begin{tabular}{c c c c c  || c c |c| c c c| c c c }        
\hline\hline                 
Run ID & Diagrams &   Area & S & E &
$w_{0}$ &$w_{a}$ & $X_{c}$  & $\alpha_{ML}$ & $log(A_{ML})$ &$\gamma_{ML}$ & $\alpha_{MT}$ & $log(A_{MT})$ &$\gamma_{MT}$\\
\hline
\hline
\small
 M10 & M(50\% error), z &   10x10,000 & & Y& -1.02$\pm 0.05$ & 0.12 $\pm 0.10$& 0.23$\pm 0.009$ & 0.51 $\pm 0.010$& 0.25 $\pm 0.015$ &0.026 $\pm 0.065$& 1.94$\pm 0.35$& 0.48 $\pm 0.06$ & 0.14 $\pm 0.09$\\
A10 & CR, HR, $r_c$, z &   10x10,000 &Y &Y& -1.01 $\pm 0.06$ & -0.06 $\pm 0.10$& 0.24 $\pm 0.003$& 0.52 $\pm 0.003$& 0.25$\pm 0.006$ & 0.009 $\pm 0.050$ &1.50 $\pm 0.010$ &0.46 $\pm 0.008$& -0.03 $\pm 0.030$\\
\hline
M12 & M(50\% error), z & 10x100 & & Y& -0.84 $\pm 0.20$ & -1.20 $\pm 1.20$ & - & -& -& -& -& -&-\\
B12 & CR, HR, $r_{c}$, z & 10x100 &Y & Y& -0.97 $\pm 0.15$ & -0.38 $\pm 0.80$ & - & -& -& -& -& -&-\\
\hline \hline
Model & & && & -1.00& 0.00& 0.240& 0.520&   0.250 & 0.00 & 1.49 & 0.460& 0.00\\
\hline \hline
\end{tabular}
\caption{Example run on the Dark Energy equation of state, where the scaling factor of the M-Rc relation and the three parameters of both the M-L and M-T relations are let free (Runs M10 and A10). The model values are those from Table \ref{clusterphysics}   (scatters are fixed to the values assumed in this Table). Runs M12 and B12 pertain to 100 \dd\ surveys, for which we assumed that the scaling relations are perfectly known.  The quoted values are the mean of the ten best-10 and the  corresponding dispersion is assumed to be representative of the 1$\sigma$ accuracy (the last digit is rounded to the next half integer).
\newline
Five first columns: as in Table \ref{runlist}
\newline
Columns 6 to 14: Fit results 
} 
\label{runA10} 
\end{sidewaystable*}

\section{Summary and conclusions}
\label{conclusion}

This paper constitutes the second step in the systematic evaluation of an X-ray observable-based procedure (ASpiX) for the cosmological analysis of cluster surveys.  The principle of the ASpiX method is to perform a  fit in a 4D observed parameter space instead of using the traditional 2D  reconstructed [M,z] plane. The four observables considered here are: count-rate, hardness ratio, angular core radius and redshift. Although this sounds like  adding more dimensions to the problem,  the modelling of the selection function is drastically simplified.   From the practical point of view, the method does not require individual cluster-mass calculations. ASpiX bypasses this tedious step that has to be iteratively performed for each trial cosmology in the bottom-up fitting approach.
This method cannot obviously rival the traditional approach involving deep-pointed X-ray observations along with ancillary data from other wavebands and, fundamentally, faces the same uncertainties as to the observable-mass transformation (except when working directly with hydrodynamical simulations - as proposed in the second-to-last bullet of the following summarised conclusions).
But because ASpiX is based on simple observable information, it allows us to include the entire detected cluster population in the cosmological analysis, even when shallow survey observations alone are available.
In order to sketch the main characteristics of this approach, we considered a simplified toy model of the cluster population for areas of 100 and 10,000 \dd\ and perform the fit using the {\sc Amoeba} minimisation code. For the time being we considered a limited set of free parameters namely, $\Omega_{m}, ~\sigma_{8},~ X_{c}, ~w_{0}, w_{a}$, possibly complemented by six more parameters describing the cluster physics, as encoded in the scaling relations. The caveat, that this particular free-parameter combination may favour the ASpiX method compared to dn/dM/dz, has been tested for by letting both the cosmological and cluster physical parameters free for the case of 10,000 \dd\ surveys.  
Our main results indicate that ASpiX is at least as efficient as the N(M,z) route; and for this, photometric redshift accuracy is sufficient (we use five bins of $\delta z = 0.2$).  ASpiX has moreover the ability to provide useful insights into latent degeneracies between cosmology and cluster physics. These results call for a number of more general comments and we outline below the upgrades foreseen to make ASpiX a fully validated cosmological tool.

\begin{itemize}

\item[•]  We neglected a number of sources of error (such as e.g. the inaccuracy of the selection function and the sample variance) but they affect  both ASpiX and the N(M,z) approach in the same way. Although adequate for the comparative goals of the present paper, the predicted uncertainties on the cosmological and cluster parameters are certain to be larger in the real world; more realistic realisations will be considered in the following articles of the series. 

\item[•] The present results are very promising, but the power of ASpiX should  be considered independently from the {\sc  Amoeba} module used to scan the likelihood space in this paper. Two approaches can be foreseen for future validation studies: (i) Either refine the calibration of the{\sc {\sc }} output and the phenomenological error budget  by considering more realistic simulated cluster catalogues or (ii) switch to  enhanced MCMC codes \cite[like {\sc Multinest,}][]{feroz09} that avoid remaining trapped in local minima and directly provide error estimates. The computational time will be an important factor for the final choice.

\item[•]  An approach similar to ASpiX has been proposed by \cite{balaguera14}, who showed that information on the cluster scaling relations can be recovered by considering the  luminosity-weighted cluster power spectrum. From a more fundamental point of view, the numerical values of the coefficients of the scaling relations are of limited interest. Notwithstanding the fact that they provide clues  on the self-similarity of clusters as a function of mass and redshift, they provide little information on the non-gravitational physics at work in the ICM.  Basically, what is relevant is: (i) the cosmological parameters, and (ii) the physics that drives cluster formation and evolution and which is responsible for the cluster X-ray properties. Scaling relations can be considered as an intermediary product and remain, as a matter of fact, a very much debated topic: low-z determinations are still being questioned \citep{giles16}  and recent results suggest that the evolution of the core does not follow that of regions beyond $r > 0.2 r_{500}$ \citep{mcdonald17}. This results from the difficulty in defining a methodology that accounts for all factors entering the determination of the SR. Because the ASpiX method makes use of scaling relations in a transparent fashion, these can be considered as the ``catalysts'' of the cosmological analysis. 

\item[•] One can envisage for a perhaps  not so distant future, the on-the-fly production of XOD directly from numerical simulations. This approach would by-pass the modelling of scaling relations as currently implemented in our analytical code (slope, normalisation, evolution and scatter). The search for the best set of parameters (cosmology + cluster physics [time-scales and amplitude of non-gravitational energy input in the ICM, ...]) would simply be achieved by comparing the observed diagram with the simulated ones - and find the one that has the highest likelihood.
We mention that a similar approach involving simulations already proved efficient for the power spectrum of the Lyman-$\alpha$ forest \cite[e.g.][]{viel06, borde14}.
In our case, the use of simulations would involve a moreover drastic simplification by discarding any consideration of the state of the clusters involved in the scaling relations, which is today a source of lengthy debate (cool-core, virialisation stage, mergers, hydrostatic bias, presence of AGN) including the way the scaling relations themselves are determined  \cite[scatter,  parametrisation of evolution, mathematical fitting; e.g.][]{pratt09}. Finally, the current work assesses that an ASpiX-inspired methodology should be the most efficient approach to deal with the eRosita cluster sample (some 100,000 clusters to be detected at a mean equivalent XMM depth of 4 ks) at least during the first phase of the survey analysis \citep{clerc12a}.

\item[•] As to the immediate future, the following papers in this series will evaluate the performances of ASpiX on data with increasing complexity in order to more accurately model real-sky observing conditions. Paper IV extends the present study to template-based N-body simulations. This allows us to switch from ideal toy-catalogues to source lists extracted by {\sc Xamin} on 25 \dd\ mosaics of simulated light-cones. We also compare the Amoeba and MCMC analyses of the XOD. Paper V will test ASpiX on hydrodynamical simulations and further investigate the impact of the accuracy of the selection function for more realistic modellings of the cluster shapes, scaling relations and AGN activity.

\end{itemize}

\begin{acknowledgements}
We thank Jean Ballet and Mauro Sereno for useful discussions. Elias Koulouridis acknowledges funding from the Centre National d'Etudes Spatiales.
\end{acknowledgements}

\bibliographystyle{aa}
\bibliography{../../../mmplib}{}

\begin{thebibliography}{36}
\expandafter\ifx\csname natexlab\endcsname\relax\def\natexlab#1{#1}\fi

\bibitem[{{Albrecht} {et~al.}(2006){Albrecht}, {Bernstein}, {Cahn}, {Freedman},
  {Hewitt}, {Hu}, {Huth}, {Kamionkowski}, {Kolb}, {Knox}, {Mather}, {Staggs},
  \& {Suntzeff}}]{albrecht06}
{Albrecht}, A., {Bernstein}, G., {Cahn}, R., {et~al.} 2006, ArXiv Astrophysics
  e-prints

\bibitem[{{Allen} {et~al.}(2011){Allen}, {Evrard}, \& {Mantz}}]{allen11}
{Allen}, S.~W., {Evrard}, A.~E., \& {Mantz}, A.~B. 2011, \araa, 49, 409

\bibitem[{{Applegate} {et~al.}(2016){Applegate}, {Mantz}, {Allen}, {der
  Linden}, {Morris}, {Hilbert}, {Kelly}, {Burke}, {Ebeling}, {Rapetti}, \&
  {Schmidt}}]{appelgate16}
{Applegate}, D.~E., {Mantz}, A., {Allen}, S.~W., {et~al.} 2016, \mnras, 457,
  1522

\bibitem[{{Arnaud} {et~al.}(2005){Arnaud}, {Pointecouteau}, \&
  {Pratt}}]{arnaud05}
{Arnaud}, M., {Pointecouteau}, E., \& {Pratt}, G.~W. 2005, \aap, 441, 893

\bibitem[{{Balaguera-Antol{\'{\i}}nez}(2014)}]{balaguera14}
{Balaguera-Antol{\'{\i}}nez}, A. 2014, \aap, 563, A141

\bibitem[{{Balaguera-Antol{\'{\i}}nez}
  {et~al.}(2011){Balaguera-Antol{\'{\i}}nez}, {S{\'a}nchez}, {B{\"o}hringer},
  {Collins}, {Guzzo}, \& {Phleps}}]{balaguera11}
{Balaguera-Antol{\'{\i}}nez}, A., {S{\'a}nchez}, A.~G., {B{\"o}hringer}, H.,
  {et~al.} 2011, \mnras, 413, 386

\bibitem[{{Borde} {et~al.}(2014){Borde}, {Palanque-Delabrouille}, {Rossi},
  {Viel}, {Bolton}, {Y{\`e}che}, {LeGoff}, \& {Rich}}]{borde14}
{Borde}, A., {Palanque-Delabrouille}, N., {Rossi}, G., {et~al.} 2014, \jcap, 7,
  005

\bibitem[{{Brax} {et~al.}(2015){Brax}, {Rizzo}, \& {Valageas}}]{brax15}
{Brax}, P., {Rizzo}, L.~A., \& {Valageas}, P. 2015, \prd, 92, 043519

\bibitem[{{Cataneo} {et~al.}(2015){Cataneo}, {Rapetti}, {Schmidt}, {Mantz},
  {Allen}, {Applegate}, {Kelly}, {von der Linden}, \& {Morris}}]{cataneo15}
{Cataneo}, M., {Rapetti}, D., {Schmidt}, F., {et~al.} 2015, \prd, 92, 044009

\bibitem[{{Chevallier} \& {Polarski}(2001)}]{chevallier01}
{Chevallier}, M. \& {Polarski}, D. 2001, International Journal of Modern
  Physics D, 10, 213

\bibitem[{{Clerc} {et~al.}(2012){Clerc}, {Pierre}, {Pacaud}, \&
  {Sadibekova}}]{clerc12a}
{Clerc}, N., {Pierre}, M., {Pacaud}, F., \& {Sadibekova}, T. 2012, \mnras, 423,
  3545

\bibitem[{{de Haan} {et~al.}(2016){de Haan}, {Benson}, {Bleem}, {Allen},
  {Applegate}, {Ashby}, {Bautz}, {Bayliss}, {Bocquet}, {Brodwin}, {Carlstrom},
  {Chang}, {Chiu}, {Cho}, {Clocchiatti}, {Crawford}, {Crites}, {Desai},
  {Dietrich}, {Dobbs}, {Doucouliagos}, {Foley}, {Forman}, {Garmire}, {George},
  {Gladders}, {Gonzalez}, {Gupta}, {Halverson}, {Hlavacek-Larrondo},
  {Hoekstra}, {Holder}, {Holzapfel}, {Hou}, {Hrubes}, {Huang}, {Jones},
  {Keisler}, {Knox}, {Lee}, {Leitch}, {von der Linden}, {Luong-Van}, {Mantz},
  {Marrone}, {McDonald}, {McMahon}, {Meyer}, {Mocanu}, {Mohr}, {Murray},
  {Padin}, {Pryke}, {Rapetti}, {Reichardt}, {Rest}, {Ruel}, {Ruhl},
  {Saliwanchik}, {Saro}, {Sayre}, {Schaffer}, {Schrabback}, {Shirokoff},
  {Song}, {Spieler}, {Stalder}, {Stanford}, {Staniszewski}, {Stark}, {Story},
  {Stubbs}, {Vanderlinde}, {Vieira}, {Vikhlinin}, {Williamson}, \&
  {Zenteno}}]{dehaan16}
{de Haan}, T., {Benson}, B.~A., {Bleem}, L.~E., {et~al.} 2016, ArXiv e-prints

\bibitem[{{Farahi} {et~al.}(2016){Farahi}, {Evrard}, {Rozo}, {Rykoff}, \&
  {Wechsler}}]{farahi16}
{Farahi}, A., {Evrard}, A.~E., {Rozo}, E., {Rykoff}, E.~S., \& {Wechsler},
  R.~H. 2016, \mnras, 460, 3900

\bibitem[{{Feroz} {et~al.}(2009){Feroz}, {Hobson}, \& {Bridges}}]{feroz09}
{Feroz}, F., {Hobson}, M.~P., \& {Bridges}, M. 2009, \mnras, 398, 1601

\bibitem[{{Giles} {et~al.}(2016){Giles}, {Maughan}, {Pacaud}, {Lieu}, {Clerc},
  {Pierre}, {Adami}, {Chiappetti}, {D{\'e}mocl{\'e}s}, {Ettori}, {Le
  F{\'e}vre}, {Ponman}, {Sadibekova}, {Smith}, {Willis}, \& {Ziparo}}]{giles16}
{Giles}, P.~A., {Maughan}, B.~J., {Pacaud}, F., {et~al.} 2016, \aap, 592, A3

\bibitem[{{Le Brun} {et~al.}(2014){Le Brun}, {McCarthy}, {Schaye}, \&
  {Ponman}}]{lebrun14}
{Le Brun}, A.~M.~C., {McCarthy}, I.~G., {Schaye}, J., \& {Ponman}, T.~J. 2014,
  \mnras, 441, 1270

\bibitem[{{Lin} {et~al.}(2003){Lin}, {Mohr}, \& {Stanford}}]{lin03}
{Lin}, Y.-T., {Mohr}, J.~J., \& {Stanford}, S.~A. 2003, \apj, 591, 749

\bibitem[{{Mantz} {et~al.}(2010){Mantz}, {Allen}, {Rapetti}, \&
  {Ebeling}}]{mantz10a}
{Mantz}, A., {Allen}, S.~W., {Rapetti}, D., \& {Ebeling}, H. 2010, \mnras, 406,
  1759

\bibitem[{{McDonald} {et~al.}(2017){McDonald}, {Allen}, {Bayliss}, {Benson},
  {Bleem}, {Brodwin}, {Bulbul}, {Carlstrom}, {Forman}, {Hlavacek-Larrondo},
  {Garmire}, {Gaspari}, {Gladders}, {Mantz}, \& {Murray}}]{mcdonald17}
{McDonald}, M., {Allen}, S.~W., {Bayliss}, M., {et~al.} 2017, ArXiv e-prints

\bibitem[{{Moretti} {et~al.}(2003){Moretti}, {Campana}, {Lazzati}, \&
  {Tagliaferri}}]{moretti03}
{Moretti}, A., {Campana}, S., {Lazzati}, D., \& {Tagliaferri}, G. 2003, \apj,
  588, 696

\bibitem[{{Nelder} \& {Mead}(1965)}]{nelder65}
{Nelder}, J.~A. \& {Mead}, R. 1965, The Computer Journal, 4, 308

\bibitem[{{Nelson} {et~al.}(2014){Nelson}, {Lau}, {Nagai}, {Rudd}, \&
  {Yu}}]{nelson14}
{Nelson}, K., {Lau}, E.~T., {Nagai}, D., {Rudd}, D.~H., \& {Yu}, L. 2014, \apj,
  782, 107

\bibitem[{{Pacaud} {et~al.}(2016){Pacaud}, {Clerc}, {Giles}, {Adami},
  {Sadibekova}, {Pierre}, {Maughan}, {Lieu}, {Le F{\`e}vre}, {Alis}, {Altieri},
  {Ardila}, {Baldry}, {Benoist}, {Birkinshaw}, {Chiappetti},
  {D{\'e}mocl{\`e}s}, {Eckert}, {Evrard}, {Faccioli}, {Gastaldello}, {Guennou},
  {Horellou}, {Iovino}, {Koulouridis}, {Le Brun}, {Lidman}, {Liske},
  {Maurogordato}, {Menanteau}, {Owers}, {Poggianti}, {Pomar{\`e}de}, {Pompei},
  {Ponman}, {Rapetti}, {Reiprich}, {Smith}, {Tuffs}, {Valageas}, {Valtchanov},
  {Willis}, \& {Ziparo}}]{pacaud16}
{Pacaud}, F., {Clerc}, N., {Giles}, P.~A., {et~al.} 2016, \aap, 592, A2

\bibitem[{{Pacaud} {et~al.}(2007){Pacaud}, {Pierre}, {Adami}, {Altieri},
  {Andreon}, {Chiappetti}, {Detal}, {Duc}, {Galaz}, {Gueguen}, {Le F{\`e}vre},
  {Hertling}, {Libbrecht}, {Melin}, {Ponman}, {Quintana}, {Refregier},
  {Sprimont}, {Surdej}, {Valtchanov}, {Willis}, {Alloin}, {Birkinshaw},
  {Bremer}, {Garcet}, {Jean}, {Jones}, {Le F{\`e}vre}, {Maccagni}, {Mazure},
  {Proust}, {R{\"o}ttgering}, \& {Trinchieri}}]{pacaud07}
{Pacaud}, F., {Pierre}, M., {Adami}, C., {et~al.} 2007, \mnras, 382, 1289

\bibitem[{{Pacaud} {et~al.}(2006){Pacaud}, {Pierre}, {Refregier}, {Gueguen},
  {Starck}, {Valtchanov}, {Read}, {Altieri}, {Chiappetti}, {Gandhi}, {Garcet},
  {Gosset}, {Ponman}, \& {Surdej}}]{pacaud06}
{Pacaud}, F., {Pierre}, M., {Refregier}, A., {et~al.} 2006, \mnras, 372, 578

\bibitem[{{Pierre} {et~al.}(2016){Pierre}, {Pacaud}, {Adami}, {Alis},
  {Altieri}, {Baran}, {Benoist}, {Birkinshaw}, {Bongiorno}, {Bremer}, {Brusa},
  {Butler}, {Ciliegi}, {Chiappetti}, {Clerc}, {Corasaniti}, {Coupon}, {De
  Breuck}, {Democles}, {Desai}, {Delhaize}, {Devriendt}, {Dubois}, {Eckert},
  {Elyiv}, {Ettori}, {Evrard}, {Faccioli}, {Farahi}, {Ferrari}, {Finet},
  {Fotopoulou}, {Fourmanoit}, {Gandhi}, {Gastaldello}, {Gastaud},
  {Georgantopoulos}, {Giles}, {Guennou}, {Guglielmo}, {Horellou}, {Husband},
  {Huynh}, {Iovino}, {Kilbinger}, {Koulouridis}, {Lavoie}, {Le Brun}, {Le
  Fevre}, {Lidman}, {Lieu}, {Lin}, {Mantz}, {Maughan}, {Maurogordato},
  {McCarthy}, {McGee}, {Melin}, {Melnyk}, {Menanteau}, {Novak}, {Paltani},
  {Plionis}, {Poggianti}, {Pomarede}, {Pompei}, {Ponman}, {Ramos-Ceja},
  {Ranalli}, {Rapetti}, {Raychaudury}, {Reiprich}, {Rottgering}, {Rozo},
  {Rykoff}, {Sadibekova}, {Santos}, {Sauvageot}, {Schimd}, {Sereno}, {Smith},
  {Smol{\v c}i{\'c}}, {Snowden}, {Spergel}, {Stanford}, {Surdej}, {Valageas},
  {Valotti}, {Valtchanov}, {Vignali}, {Willis}, \& {Ziparo}}]{pierre16}
{Pierre}, M., {Pacaud}, F., {Adami}, C., {et~al.} 2016, \aap, 592, A1

\bibitem[{{Pierre} {et~al.}(2011){Pierre}, {Pacaud}, {Juin}, {Melin},
  {Valageas}, {Clerc}, \& {Corasaniti}}]{pierre11}
{Pierre}, M., {Pacaud}, F., {Juin}, J.~B., {et~al.} 2011, \mnras, 414, 1732

\bibitem[{{Pillepich} {et~al.}(2012){Pillepich}, {Porciani}, \&
  {Reiprich}}]{pillepich12}
{Pillepich}, A., {Porciani}, C., \& {Reiprich}, T.~H. 2012, \mnras, 422, 44

\bibitem[{{Planck Collaboration} {et~al.}(2015){Planck Collaboration}, {Ade},
  {Aghanim}, {Arnaud}, {Ashdown}, {Aumont}, {Baccigalupi}, {Banday},
  {Barreiro}, {Bartlett}, \& et~al.}]{planck2015XXIV}
{Planck Collaboration}, {Ade}, P.~A.~R., {Aghanim}, N., {et~al.} 2015, ArXiv
  e-prints

\bibitem[{{Pratt} {et~al.}(2009){Pratt}, {Croston}, {Arnaud}, \&
  {B{\"o}hringer}}]{pratt09}
{Pratt}, G.~W., {Croston}, J.~H., {Arnaud}, M., \& {B{\"o}hringer}, H. 2009,
  \aap, 498, 361

\bibitem[{{Rozo} {et~al.}(2015){Rozo}, {Rykoff}, {Bartlett}, \&
  {Melin}}]{rozo15}
{Rozo}, E., {Rykoff}, E.~S., {Bartlett}, J.~G., \& {Melin}, J.-B. 2015, \mnras,
  450, 592

\bibitem[{{Snowden} {et~al.}(2008){Snowden}, {Mushotzky}, {Kuntz}, \&
  {Davis}}]{snowden08}
{Snowden}, S.~L., {Mushotzky}, R.~F., {Kuntz}, K.~D., \& {Davis}, D.~S. 2008,
  \aap, 478, 615

\bibitem[{{Tinker} {et~al.}(2008){Tinker}, {Kravtsov}, {Klypin}, {Abazajian},
  {Warren}, {Yepes}, {Gottl{\"o}ber}, \& {Holz}}]{tinker08}
{Tinker}, J., {Kravtsov}, A.~V., {Klypin}, A., {et~al.} 2008, \apj, 688, 709

\bibitem[{{Veropalumbo} {et~al.}(2016){Veropalumbo}, {Marulli}, {Moscardini},
  {Moresco}, \& {Cimatti}}]{veropalumbo16}
{Veropalumbo}, A., {Marulli}, F., {Moscardini}, L., {Moresco}, M., \&
  {Cimatti}, A. 2016, \mnras, 458, 1909

\bibitem[{{Viel} \& {Haehnelt}(2006)}]{viel06}
{Viel}, M. \& {Haehnelt}, M.~G. 2006, \mnras, 365, 231

\bibitem[{{Vikhlinin} {et~al.}(2009){Vikhlinin}, {Kravtsov}, {Burenin},
  {Ebeling}, {Forman}, {Hornstrup}, {Jones}, {Murray}, {Nagai}, {Quintana}, \&
  {Voevodkin}}]{vikhlinin09}
{Vikhlinin}, A., {Kravtsov}, A.~V., {Burenin}, R.~A., {et~al.} 2009, \apj, 692,
  1060

\end{thebibliography}

\appendix
\section{ASpiX output}
\label{output}
We present a compilation of the ASpiX outputs. Each configuration listed in Tables \ref{runlist} \& \ref{runA10} is explored by running 100 times ASpiX with random starting values for the fitted parameters in question. For each configuration we display:
\begin{itemize}
\item[•] A table with (i) either the results from the best, averaged best-10, or best-20 likelihood trajectories when processing a single  10,000 \dd\ catalogue (ii) or  the  averaged best, best-10  and all likelihoods for the ten  distinct 100 and 10,000 \dd\ catalogues. The second column then gives the 1$\sigma$ dispersion around the mean. 
\item[•]
The histograms of the output values from all converged individual {\sc Amoeba} runs. The red vertical line stands for the fiducial value and the two blue lines for $\pm 5 \%$ off the fiducial. For the 10$\times$100 and 10$\times$10,000 \dd\ runs, the pink error-bar indicates the 1$\sigma$ dispersion of the values given by the best likelihood of each catalogue and the yellow bar is when considering all outputs; the black crosses  show the individual parameter values given by the best likelihoods.
\end{itemize}

\clearpage

\section{Exploring degeneracies}
\label{degeneracy}

In this Appendix, we show two examples of degenerate parameter sets present in either the XOD or in the mass representation. We show under which conditions, the degeneracy can be broken. \\

\begin{figure}
   \centering
\includegraphics[width=8cm]{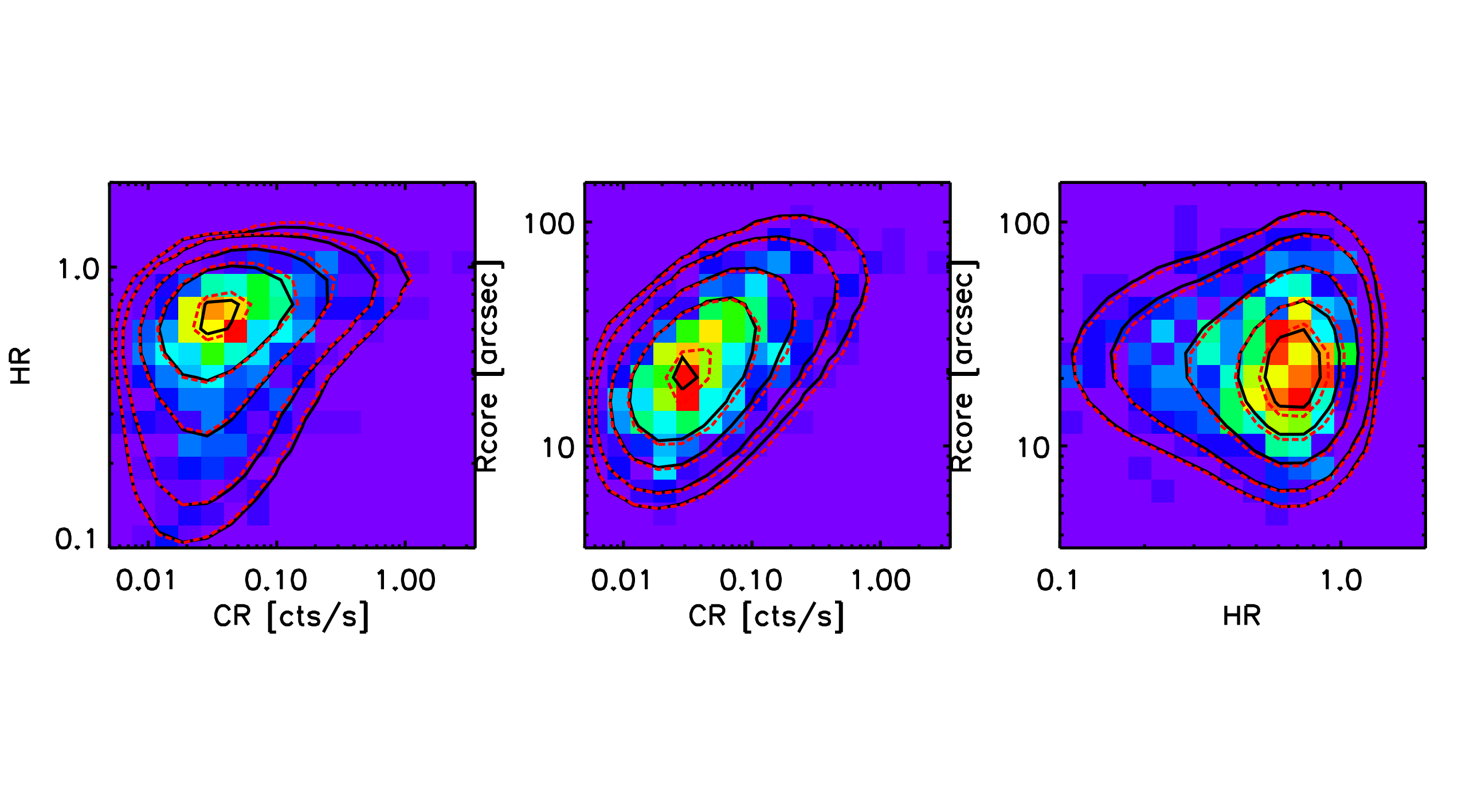}
   \caption{Example of two different models yielding extremely similar integrated HR-CR-$r_{c}$ diagrams. Black contours: the fiducial model, $\Omega_{m} =0.23,~ \sigma_{8}=0.83,~ X_{co}=0.24, w_{0}=-1$. Red contours: $\Omega_{m} =0.295, ~\sigma_{8}=0.768,~ X_{c}=0.203, ~w_{0}=-2.251$ (Best LH). The background pixel image shows one catalogue realisation for 100 \dd . For the sake of clarity, error measurements are included neither in this Figure, nor in the following two (but scatter is implemented in the three scaling relations). }
\label{B2degenerate}
\end{figure}

\begin{figure}
   \centering
\includegraphics[width=4cm]{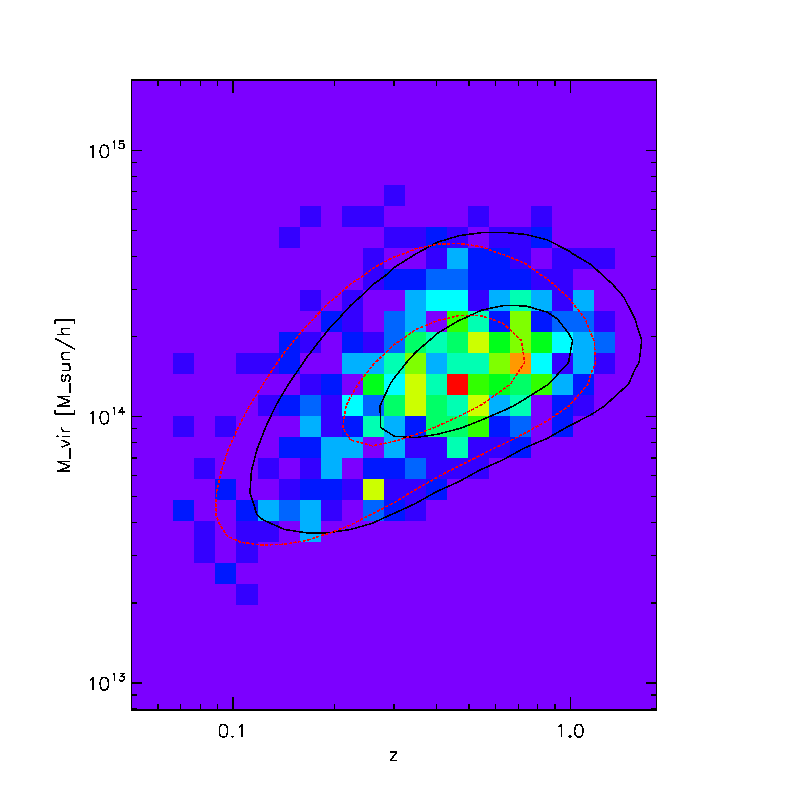}
   \caption{Mass-redshift distribution corresponding to the catalogue of Fig. \ref{B2degenerate}; same model contours.}
\label{B2degenerate-mass}
\end{figure}

\begin{figure}
   \centering
\includegraphics[width=8cm]{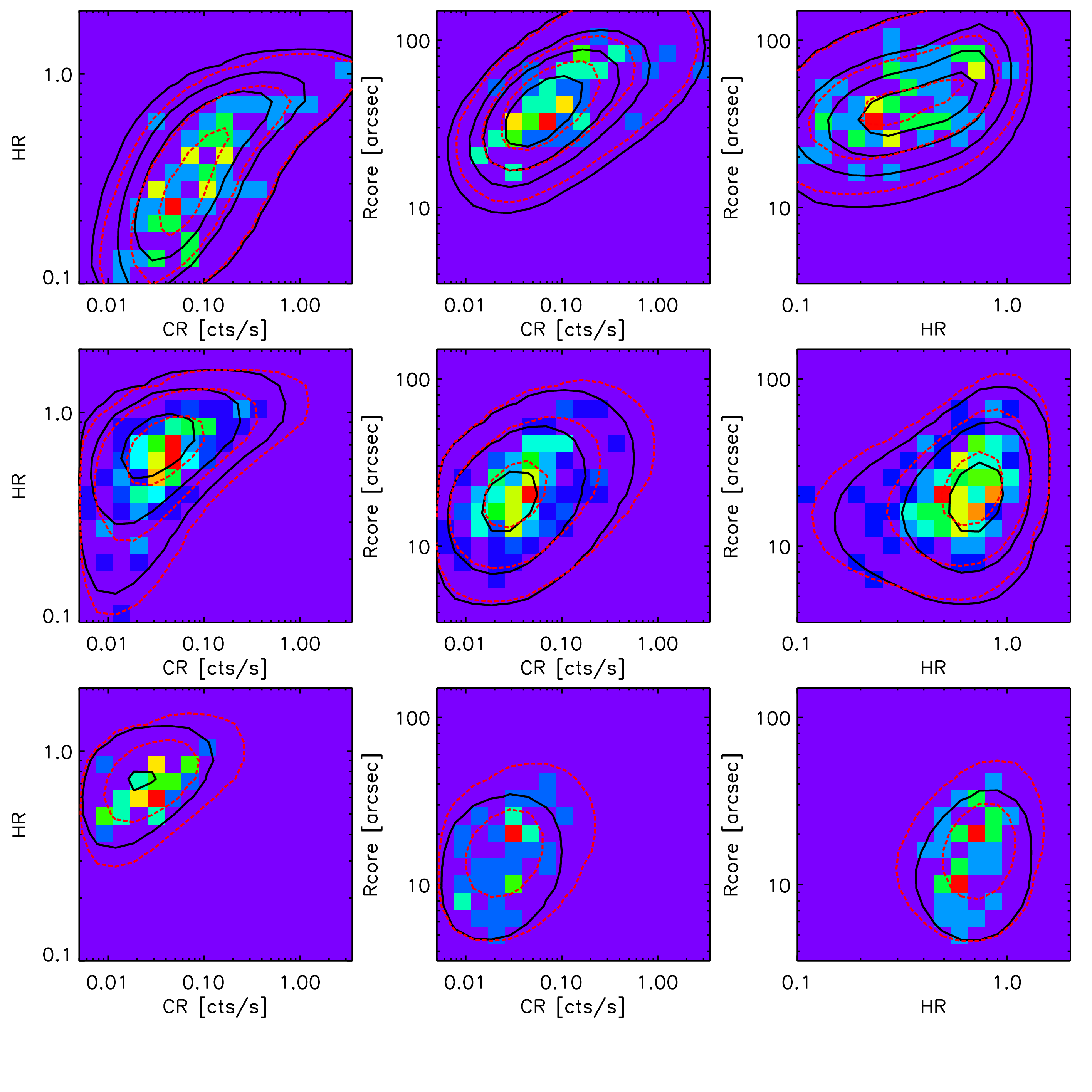}   
   \caption{Same models as in Fig. \ref{B2degenerate}, but this time decomposed into three redshift slices at $z = 0.1, 0.5, 1$. The degeneracy is broken and ASpiX finds a solution quite close to the fiducial model for this particular 100 \dd\ catalogue realisation, namely: $\Omega_{m} = 0.236, ~\sigma_{8}=0.815,~ X_{c}=0.246, ~w_{0}=-1.02$. }
\label{B3degenerate}
\end{figure}

\noindent
{\bf Example 1: Degeneracy in the absence of redshift information}\\
One of the 100 \dd\ catalogues+ (cat-6) was found to yield the following set of parameters for the B2 configuration (no redshift information): 
$\Omega_{m} =0.295, ~\sigma_{8}=0.768,~ X_{c}=0.203, ~w_{0}=-2.251$.
The corresponding model is plotted in Fig. \ref{B2degenerate} along with the fiducial model: both models are almost indistinguishable. The mass distribution, however, clearly differentiates the two models (Fig. \ref{B2degenerate-mass}). Decomposing the XOD into redshift slices allows the ambiguity to be removed and provides a solution approaching the fiducial model to better than 3\% for $w_{0}$, namely: $\Omega_{m} = 0.236, ~\sigma_{8}=0.815,~ X_{c}=0.246, ~w_{0}=-1.02$ for the best LH,  (Fig. \ref{B3degenerate}). The fitting of the mass distribution (Fig. \ref{B2degenerate-mass}) yields: $\Omega_{m} = 0.251, ~\sigma_{8}=0.812,~ X_{c}=0.247, ~w_{0}=-1.17$.

\clearpage

\begin{figure}
   \centering
\includegraphics[width=8cm]{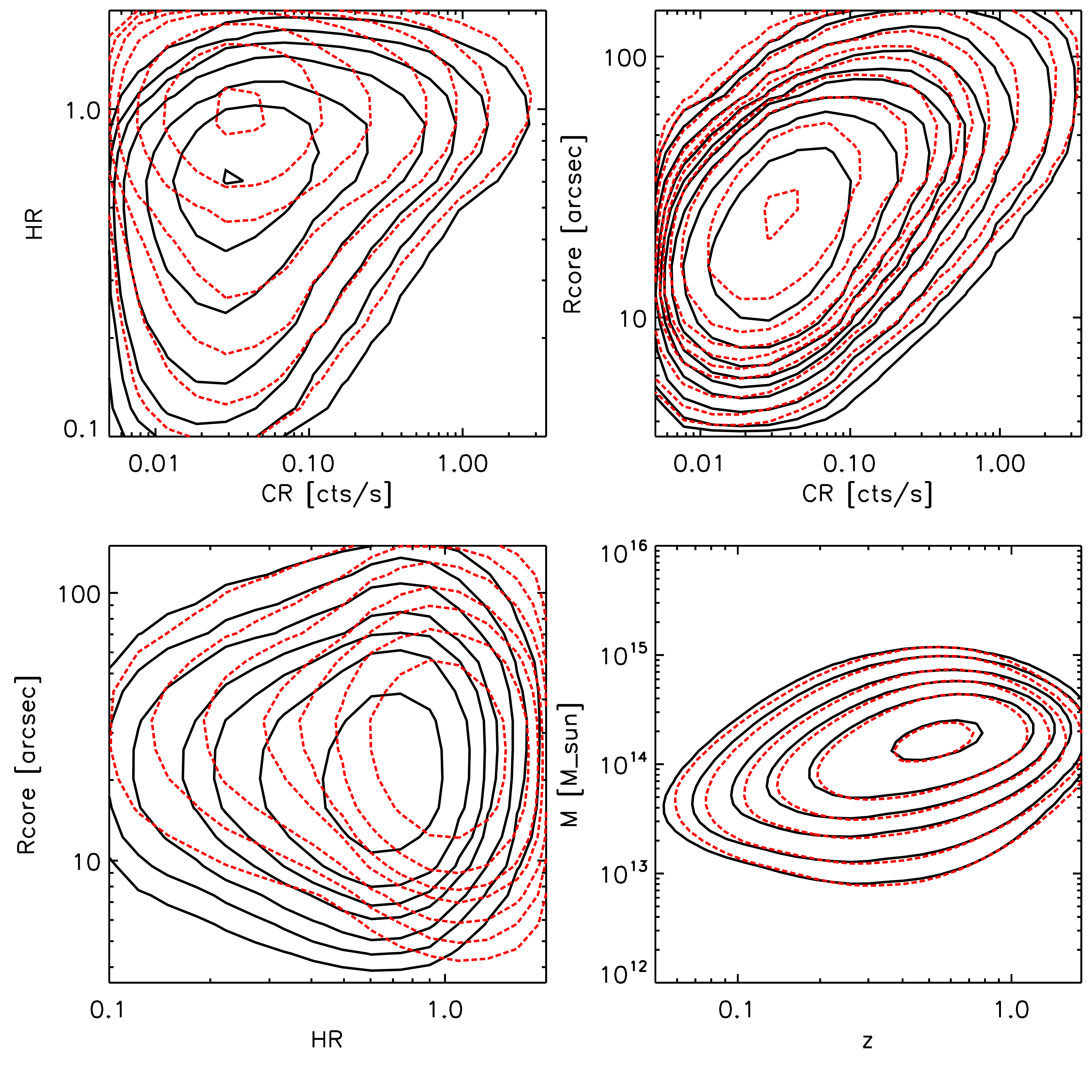}
   \caption{Illustration of a degeneracy case in the N(M,z) space which is resolved in the 3D XOD space. The black contours stand for the fiducial model ($X_{c} = 0.24, ~\gamma_{MT}= 0, \gamma_{ML} = 0, ~ w_{0}=-1, w_{a}=0$) and the red ones for the following parameter set: $X_{c} = 0.29, ~\gamma_{MT}=-2.30, \gamma_{ML}-0.54, ~ w_{0}=-0.11, w_{a}=-6.97$ .  The first  panels show the three planes of the CR-HR-$r_{c}$ space and the bottom right panel, the corresponding mass distribution.
Scatter in the scaling relations and error measurements are included (a very similar situation is  observed, when not included).}
\label{M7degenerate}
 \end{figure}

\noindent
{\bf Example 2: Degeneracy in the mass distribution}\\
For a specific configuration of the free parameter set (not mentioned in Table \ref{runlist}), one 100 \dd\ catalogue realisation (cat-3) was found to converge towards a very `exotic' solution namely:
$X_{c} = 0.29, ~\gamma_{MT}=-2.30, \gamma_{ML}=-0.54, ~ w_{0}=-0.11, w_{a}=-6.97$ with 568 objects (the fiducial model gives 577).
After investigation, the analytical model corresponding to this parameter set appears to be almost entirely degenerate with the fiducial one for the mass distribution, but not in the 3D XOD representation (Fig. \ref{M7degenerate}); this is one more indication that ASpiX  outperforms N(M,z) for the dark energy equation of state. It is even more interesting to point out that the XOD can easily discriminate the two parameter sets without the redshift information. \\

\end{document}